\documentclass[lettersize,journal]{IEEEtran}
\usepackage{amsmath,amsfonts}
\usepackage{algorithmic}
\usepackage{algorithm}
\usepackage{array}
\usepackage[caption=false,font=normalsize]{subfig}
\usepackage{textcomp}
\usepackage{stfloats}
\usepackage{url}
\usepackage{verbatim}
\usepackage{graphicx}
\usepackage{cite}
\usepackage{epsfig}
\usepackage[hidelinks]{hyperref}
\usepackage{wrapfig}
\usepackage{multirow}
\usepackage{threeparttable}

\usepackage[dvipsnames]{xcolor}
\usepackage{amsthm}
\usepackage{booktabs}
\usepackage{epstopdf}
\usepackage{dsfont}
\usepackage{soul}
\usepackage{cleveref}

\newtheorem{rmk}{Remark}

\begin{document}

\title{Deep Neural Koopman Operator-based Economic Model Predictive Control of Shipboard Carbon Capture System}

\author{Minghao Han, and 
Xunyuan Yin
\thanks{
This research is supported by the National Research Foundation, Singapore, and PUB, Singapore’s
National Water Agency under its RIE2025 Urban Solutions and Sustainability
(USS) (Water) Centre of Excellence (CoE) Programme, awarded to Nanyang
Environment \& Water Research Institute (NEWRI), Nanyang Technological
University, Singapore (NTU). This research is also supported by the Ministry of Education, Singapore, under its Academic Research Fund Tier 1 (RG63/22). Any opinions, findings and conclusions or recommendations expressed in this material are those of the author(s) and do not reflect the views of the National Research Foundation, Singapore, and PUB, Singapore’s National Water Agency.  (\emph{Corresponding author: Xunyuan Yin.}) 

Minghao Han and Xunyuan Yin are with the School of Chemistry, Chemical Engineering and Biotechnology, Nanyang Technological University, Singapore, 637459. They are also with Environmental Process Modelling Centre, Nanyang Environment and Water Research Institute (NEWRI), Nanyang Technological University, Singapore, 637141  (e-mail: minghao.han@ntu.edu.sg; xunyuan.yin@ntu.edu.sg). 

}
}

\markboth{}%
{Shell \MakeLowercase{\textit{et al.}}: Deep Neural Koopman Operator-based Economic Model Predictive Control of Shipboard Carbon Capture System}


\maketitle

\begin{abstract}
Shipboard carbon capture is a promising solution to help reduce carbon emissions in international shipping. In this work, we propose a data-driven dynamic modeling and economic predictive control approach within the Koopman framework. This integrated modeling and control approach is used to achieve safe and energy-efficient process operation of shipboard post-combustion carbon capture plants.
Specifically, we propose a deep neural Koopman operator modeling approach, based on which a Koopman model with time-varying model parameters is established. This Koopman model predicts the overall economic operational cost and key system outputs, based on accessible partial state measurements. By leveraging this learned model, {\color{black}a constrained economic predictive control scheme is developed. Despite time-varying parameters involved in the formulated model, the formulated optimization problem associated with the economic predictive control design is convex, and it can be solved efficiently during online control implementations. Extensive tests are conducted on a high-fidelity simulation environment for shipboard post-combustion carbon capture processes. Four ship operational conditions are taken into account.} The results show that the proposed method significantly improves the overall economic operational performance and carbon capture rate. Additionally, the proposed method guarantees safe operation by ensuring that hard constraints on the system outputs are satisfied.
\end{abstract}

\section{Introduction}

{\color{black}
Modern transportation is a major factor in climate change and global warming. In 2022,  3$\%$ of global greenhouse gas emissions were produced by international shipping \cite{bach2023imo}. 
It is predicted that CO$_2$ emissions from shipping could reach 1.6 billion tons by 2050 \cite{capros2013eu}. As a countermeasure, the International Maritime Organization (IMO) updated its greenhouse gas (GHG) mitigation strategy in 2023 \cite{IMO2016}. The updated strategy aims at achieving net-zero GHG emissions from global shipping by 2050, with cutting total annual GHG emissions by 70$\%$ by 2040 as an intermediate target  \cite{mepc20232023}.}


One of the promising approaches to reducing emissions from ships is the implementation of post-combustion carbon capture (PCC) systems \cite{feenstra2019ship, luo2017study}. PCC systems can be retrofitted into existing systems with minimal modifications \cite{aakko2023reduction}. Several studies have explored PCC for ships \cite{feenstra2019ship, luo2017study, ros2022advancements, bayramouglu2023application,vo2024advanced}. Ship engines produce flue gases during fuel combustion for propulsion. The flue gases are subsequently treated by a shipboard PCC plant before being released into the atmosphere. However, regenerating the captured CO$_2$ within the PCC process requires substantial heat energy \cite{decardi2018improving}. 

{\color{black}
Balancing safety, energy efficiency and the need to sustain high carbon capture performance is challenging due to the complex structure and large scale of an integrated shipboard PCC plant. Advanced control solutions are essential to ensure efficient dynamic operation. Model predictive control (MPC), as one of the most effective advanced control methodologies \cite{rawlings2000tutorial,lorenzen2016constraint,mayne2000constrained,TAC2022learningMPC,di2013stochastic,panahi2012economically}, holds the promise for optimally managing the operation of shipboard PCC processes. Based on the equilibrium dynamic model, linear MPC showed superior performance compared to a proportional-integral-derivative controller on the PCC for a 550 MWe post-combustion, supercritical, pulverized coal plant \cite{zhang2016development}.
\cite{zhang2018nonlinear} further showed that the nonlinear MPC method based on a nonlinear additive autoregressive with exogenous variables model outperformed a linear MPC design in \cite{zhang2016development} on a PCC process. 
Similar observations were made in \cite{akinola2020nonlinear}. 
\cite{patron2022integrated} integrated nonlinear MPC with moving horizon estimation and real-time optimization to achieve the economic operation of the PCC plant with only access to partial and noisy measurements.} 
Economic model predictive control (EMPC) has the ability to further reduce operational costs compared to set-point tracking MPC \cite{ellis2014tutorial,wang2018economic,decardi2018improving}. 
Particularly, the integration of EMPC with machine learning has shown great potential in improving the overall operational/production performance of complex industrial systems \cite{krishnamoorthy2021adaptive,jia2021multi,zhang2023modeling}. 
However, the significant differences in process designs, component sizes, and heat supply methods between land-based and shipboard applications \cite{luo2017study} pose challenges in transferring these control methodologies to shipboard settings.
\textcolor{black}{There have been limited results on the dynamic modeling and control of shipboard post-combustion carbon capture plants. Recently, in \cite{zhang2025machine}, a carbon capture process design was presented for shipboard applications. Building on this, a machine learning-based hybrid modeling approach and a nonlinear EMPC scheme were proposed to address scenarios where only an imperfect first-principles model is available \cite{zhang2025machine}. However, this method relies on physical knowledge of the PCC process, and the control scheme suffers from relatively complex computations \cite{zhang2025machine}.}


\textcolor{black}{Koopman operator theory has recently gained considerable attention, owing to its capability to establish linear representations for nonlinear dynamical processes \cite{koopman1931hamiltonian}. This mathematical framework essentially provides a global linear embedding mechanism for nonlinear systems, thereby enabling the application of linear control methodologies to analyze and control nonlinear systems \cite{korda2018linear}. However, a major challenge persists in practical implementations, that is, the theoretical formulation of Koopman operators typically resides in infinite-dimensional function spaces and can result in finite-dimensional approximations, which are not suitable from an application perspective.
To address this dimensionality challenge, building approximated Koopman operators using system operational data has been considered a promising alternative. Accordingly, computationally efficient data-driven Koopman approximation algorithms, including dynamic mode decomposition (DMD) \cite{korda2020optimal} and its enhanced variant extended dynamic mode decomposition (EDMD) \cite{folkestad2020extended} have been proposed and widely adopted. Both DMD and EDMD employ least-squares optimization to derive linear Koopman operator approximations, while EDMD extends the original DMD with nonlinear lifting functions.
The implementation of EDMD typically requires a set of user-specified lifting functions to construct appropriate observable spaces \cite{korda2018linear}. To address this limitation, recent advancements have attempted to incorporate deep learning into the Koopman framework, and various machine learning-based Koopman modeling approaches were proposed, see, e.g., \cite{yeung2019learning,han2020deep,ping2021deep,han2023robust,li2024machine}. 
MPC has been integrated with Koopman modeling for efficient control of various nonlinear systems \cite{vsvec2023predictive,narasingam2023data,chen2024incorporating}.}
More results relevant to Koopman-based modeling and control can be found in \cite{zhang2023reduced,han2023robust_tii,chen2024deep,susuki2024control}.
Within the Koopman framework, EMPC can also be more advantageous than tracking MPC for certain nonlinear industrial processes such as PCC systems, when economic and sustainable operation is a priority. \textcolor{black}{In \cite{albalawi2023koopman}, EDMD was integrated with EMPC for the economic operation of a chemical process. Similarly, in \cite{mayfrank2024end}, a Koopman operator model was trained within a reinforcement learning framework to develop economic controllers.} In \cite{han2024efficient}, deep learning-based Koopman modeling was leveraged to develop a convex input-output EMPC for the efficient operation of water treatment facilities. However, this approach lacks the ability to predict key output variables of the process and does not effectively handle hard operational constraints.

In this work, we propose a learning-based Koopman economic predictive control framework for the safe and efficient operation of shipboard PCC systems. Specifically, we consider the shipboard carbon capture design presented in \cite{zhang2025machine}, and we introduce a learning-based Koopman modeling framework called the deep neural Koopman operator (DNKO) model. The Koopman model predicts future economic operational costs and key outputs based on partial state measurements and system inputs. Some model parameters are dynamically adjusted using trained neural networks to improve prediction accuracy. By leveraging the input and economic cost data, the develooped DNKO model learns a computation-efficient latent representation, and it predicts future critical performance indicators for shipboard PCC operations. Building on the DNKO model, we formulate a computationally efficient convex EMPC design. This Koopman-based EMPC control method is applied to the shipboard PCC system, and it significantly improves economic operational performance, safety, and carbon capture rate across \textcolor{black}{four different operational conditions.}
The contributions of this work are summarized as follows:
\begin{itemize}
    \item We propose a learning-based Koopman modeling approach to predict future economic costs and key outputs of the shipboard PCC system, alleviating the need for full-state information.
    \item We formulate a Koopman-based constrained economic predictive control scheme to ensure the safe and economical real-time operation of the highly nonlinear shipboard PCC system.
    \item As compared to competitive baseline methods, \textcolor{black}{our proposed framework provides improved economical operational performance across different operational conditions, with the highest carbon capture rate up to 87.12$\%$}.
\end{itemize}

The remainder is organized as follows. Section~\ref{sec:system} provides a comprehensive overview of the shipboard PCC system and outlines the learning and control problem. In Section~\ref{sec:DNKO}, we introduce the building blocks of the proposed DNKO model approach and model training. Section~\ref{sec:empc} presents the proposed constrained economic predictive control method. The modeling and control results, along with the performance comparison against baselines, are presented in Section~\ref{sec:results}. Finally, conclusions and discussions are presented in Section~\ref{sec:conclusion}.

\section{Shipboard Carbon Capture System and Problem Formulation}\label{sec:system}

\begin{figure*}
    \centering
    \includegraphics[width=0.8\textwidth]{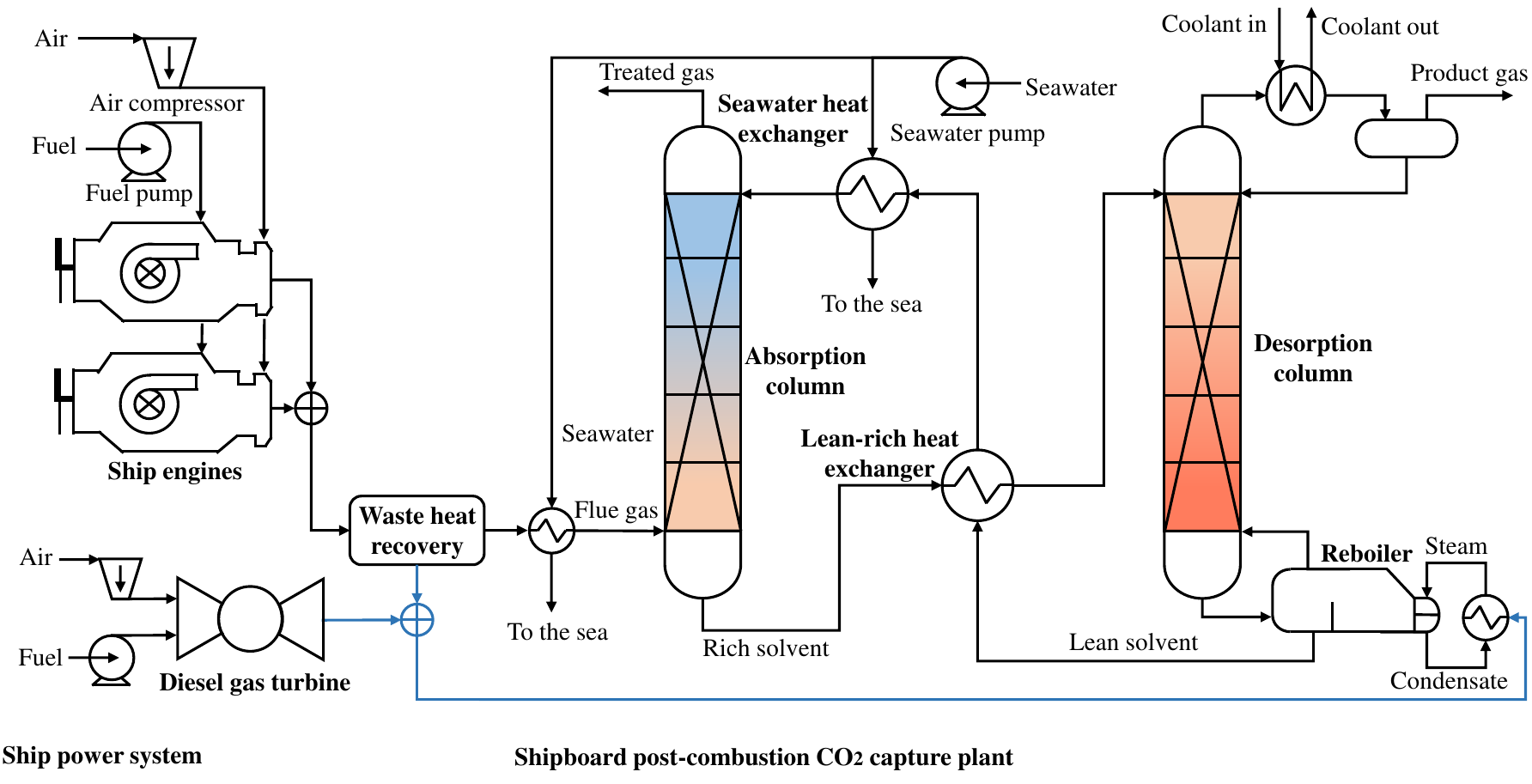}
    \caption{A schematic view of the integrated shipboard PCC system {\color{black}\cite{zhang2025machine}}. 
    }
    \label{fig:PCCS}
\end{figure*}

We consider a shipboard carbon capture design presented in \cite{zhang2025machine}. 
The shipboard carbon capture process consists of two primary components: the ship engine system, and the shipboard post-combustion carbon capture (PCC) plant \cite{zhang2025machine}. A schematic of this integrated process is provided in Fig. \ref{fig:PCCS}. The ship engine system includes two main engines that generate propulsion power. The combustion of diesel fuel and air in these engines produces hot flue gas, from which heat energy is recovered via a waste heat recovery (WHR) system. Additional heat energy is generated by burning extra diesel fuel in a gas turbine. The heat energy from the WHR system and the gas turbine is directly supplied to the reboiler.
The absorption column plays a critical role in capturing CO$_2$. In the desorption column, the CO$_2$ captured in the solvent is released through thermal regeneration, using the heat energy supplied by the reboiler. After regeneration, the hot lean solvent is cooled in lean-rich solvent and seawater heat exchangers before entering the absorption column \cite{zhang2025machine}. \textcolor{black}{A detailed description of the shipboard PCC system can be found in \cite{zhang2025machine}. }




\subsection{Ship Engine System}

In this work, we consider a mid-size cargo ship powered by two 9L46 marine diesel engines \cite{wartsila46f}. 
In the fuel injection system, a fuel pump is used to regulate the fuel flow rate. Inside the engine, diesel fuel is mixed with high-pressure air and it combusts. The hot exhaust gases generated are processed by a waste heat recovery system to extract heat energy. The extracted heat energy is used to partially power the reboiler in the shipboard PCC process. The cooled exhaust gas is subsequently directed to the absorption column for carbon capture.

{\color{black}Various operations, including berthing, cruising, and maneuvering, can lead to varying engine loads and PCC demands \cite{winnes2010emissions}.} Particularly, as the operational condition varies during a voyage, the engine load (i.e., $\varphi_E$) fluctuates, which can significantly affect the flue gas flow rate and the dynamic behaviors of the shipboard PCC \cite{zhang2025machine}. In this work, we consider and simulate fluctuations in the engine load under different operational conditions.

\subsubsection{Flue gas flow rate}
\textcolor{black}{The flow rate of $\text{CO}_2$ contained in the flue gas is computed based on the power output of ship engines \cite{fan2022review} and the fuel consumption of the diesel gas turbine}:
\begin{equation}\label{eq:eng:co2}
\begin{aligned}
\tilde{F}_{flue,\text{CO}_2} &= \frac{r_{\text{CO}_2}}{3600 r_\text{C}} q_{fuel,\text{C}}\varphi_E 2 Q_E W_{SFOC} \\ &~~~~~ \textcolor{black}{+ \frac{r_{\text{CO}_2}}{r_\text{C}}q_{fuel,\text{C}}\tilde{F}_{fuel}}
\end{aligned}
\end{equation}
where $\tilde{F}_{flue, \text{CO}_2}$ denotes the mass flow rate of CO$_2$ contained in the flue gas in kg/s; $r_\text{C}$ and $r_{\text{CO}_2}$ represent the molar masses of carbon and CO$_2$ in kg/kmol; $q_{fuel, C}$ represents the mole fraction of carbon in the fuel; $Q_E$ is the maximum power output from one single engine in kW; $W_{SFOC}$ is fuel consumption in kg/kWh; \textcolor{black}{$\tilde{F}_{fuel}$ is the mass flow rate of fuel in kg/s}. 

{\color{black}The shipboard PCC plant in \cite{zhang2025machine} accounts only for the flue gas produced by the ship engines. In this study, we consider a broader range of fuel flow rates for the diesel gas turbine. Accordingly, as is different from\cite{zhang2025machine}, the flue gas generated by the diesel gas turbine is jointly considered with the flue gas generated by the ship engines. Therefore, (\ref{eq:eng:co2}) for calculating the CO$_2$ mass flow rate is slightly different from the formula adopted in \cite{zhang2025machine}. }

The flue gas flow rate (denoted by \(F_G\)) in m$^3$/s can be calculated following \textcolor{black}{\cite{harun2012dynamic}}:
\begin{equation}\label{eq:eng:FG}
    F_G = \frac{\tilde{F}_{flue,\text{CO}_2}}{q_{flue,\text{CO}_2}\tilde{\rho}_{flue}} 
\end{equation}
where $q_{flue,\text{CO}_2}$ represents the mole fraction of CO$_2$ contained in the flue gas, and $\tilde{\rho}_{flue}$ represents the density of the flue gas in kg/m$^3$.

\subsubsection{Waste heat recovery of flue gas}

The flue gas emitted from ship engines reaches temperatures up to approximately 360$^\circ$C. A waste heat recovery system cools the flue gas down to approximately 150$^\circ$C, and it captures the excess heat to power the reboiler of the PCC plant. This process helps reduce the energy demands on the diesel gas turbine. Subsequently, the flue gas is further cooled to 40$^\circ$C via a heat exchanger before entering the absorber.

The available waste heat energy can be calculated using the following formula \cite{jouhara2018waste}:
\begin{equation}\label{heatrec:1}
     Q_{rec} = \tilde{\rho}_{flue}  \tilde{C}_{p,flue} F_{G} (T_{rec,in} - T_{rec,out})
\end{equation}
where $Q_{rec}$ represents the recovered heat energy in kW; $\tilde{C}_{p,flue}$ is the specific heat capacity of the flue gas in kJ/(kg$\cdot$K); $T_{rec,in}$ and $T_{rec,out}$ are the initial and final temperatures of the gas in the waste heat recovery system (in Kelvin), respectively. The recovered heat $Q_{rec}$ is then used for the reboiler.

\subsubsection{Diesel gas turbine}

The diesel gas turbine burns diesel fuel to generate additional energy, which is supplied to the reboiler. This energy is utilized to produce hot steam that heats the solvent, which facilitates CO$_2$ separation. \textcolor{black}{The heat energy generated by the diesel gas turbine is calculated as \cite{danov2004modeling}}:
\begin{equation}\label{reb:steam:heat2} 
Q_{turbine} = \eta_{fuel} \tilde{F}_{fuel} \frac{\hat{H}_{steam}-\hat{H}_{water}}{\hat{H}_{steam}},
\end{equation}
where $\eta_{fuel}$ represents the calorific value of the diesel fuel in kJ/kg; $Q_{turbine}$ denotes the heat energy produced by the turbine in kW; $\hat{H}_{steam}$ and $\hat{H}_{water}$ are the specific enthalpies of saturated steam and water, respectively, at a steam pressure of 6 barG, in kJ/kg. {\color{black}The fuel flow rate, represented by $\tilde{F}_{fuel}$, can be adjusted to regulate the reboiler temperature \cite{zhang2025machine}.}

\subsection{Shipboard post-combustion carbon capture process}

Fig.~\ref{fig:PCCS} illustrates the shipboard PCC plant, which comprises five main components: the absorption column, the desorption column, the reboiler, the lean-rich solvent heat exchanger, and the seawater heat exchanger. In the absorption column, lean solvent interacts with the flue gas emitted from the ship engines, to selectively absorb CO$_2$ and separate it from other gases. The processed gas, with a reduced CO$_2$ concentration, is released into the atmosphere from the top of the absorption column.
The CO$_2$-enriched solvent is then transferred to the desorption column, in which it undergoes thermal regeneration and releases concentrated CO$_2$ gas from the top. The solvent, depleted of CO$_2$, flows from the bottom of the desorption column to the reboiler, where it is heated to approximately 120$^\circ$C to release any remaining CO$_2$, which is recycled back to the desorption column. The heated lean solvent is subsequently cooled in the lean-rich solvent heat exchanger and the seawater heat exchanger before being recycled to the top of the absorption column.

\subsubsection{Absorption column and desorption column}

The absorption column and the desorption column have similar model formulations yet differ in reaction directions. Additionally, the desorption column includes a reboiler and a condenser.

{\color{black}The dynamic equations that describe the behavior of the state variables in the two columns, derived from mass and energy balances, are adopted from \cite{harun2012dynamic,decardi2018improving}. For brevity, these equations are not included in this paper. In these two columns, the state variables include $C_{L,i}$ and $C_{G,i}$ that denote the concentrations of substance $i$ (where $i$ represents CO$_2$, MEA, H$_2$O, or N$_2$) in liquid phase and gas phase in kmol/m$^3$.}


\subsubsection{Heat exchanger}
The heat exchangers of the solvent are crucial components in the PCC process. Since there is no mass transfer between the streams and no accumulation is assumed, the heat exchanger model does not include mass balances. In this shipboard PCC plant, there are two heat exchangers: the lean-rich heat exchanger and the seawater heat exchanger. The mathematical model for the lean-rich solvent heat exchanger is based on \cite{harun2012dynamic,decardi2018improving}.

The seawater heat exchanger is used to cool the hot lean solvent exiting from the lean-rich heat exchanger further using seawater. Seawater is an efficient cooling source for the shipboard PCC process. A seawater pump extracts and regulates the flow rate of seawater directed to the heat exchanger \cite{zhang2025machine}.

The energy balance for the seawater heat exchanger is adapted from \cite{decardi2018improving, harun2012dynamic}:
\begin{equation}\label{seahe:T}
    T_{sol, out} = T_{sol, in} + \frac{\hat{\rho}_{sw} F_{sw} \tilde{C}_{p,sw}}{\hat{\rho}_{sol} F_L \tilde{C}_{p,sol}} (T_{sw,in} - T_{sw,out})
\end{equation}
where $F_{sw}$ is the seawater volumetric flow rate in m$^3$/s; $\tilde{C}_{p}$ is the heat capacity in kJ/(kg$\cdot$K); $\hat{\rho}$ represents the density of the component in kg/m$^3$. Subscripts $sw$ and $sol$ refer to seawater and solvent, respectively, while $in$ and $out$ denote the inlet and outlet of the seawater heat exchanger, respectively. It is assumed that the densities of seawater and solvent are identical. The seawater volumetric flow rate, $F_{sw}$, is a manipulated variable used to control the solvent temperature.

\subsubsection{Reboiler}

The reboiler provides heat to the bottom of the desorption column to regenerate the solvent solution. Regenerating a CO$_2$-rich amine solution requires substantial energy \cite{metz2005ipcc}. 
The solvent with a low CO$_2$ concentration exits the reboiler as a lean solvent, which is recycled for further utilization. After passing through the lean-rich heat exchanger and the seawater heat exchanger, the lean solvent returns to the absorption column. To prevent thermal degradation of MEA, in the current study, we require that the reboiler temperature does not exceed 125$^\circ$C.

\subsubsection{Mass balance}

We assume constant liquid levels and pressure in the reboiler. Consequently, the mass balance equation for the reboiler is given as \textcolor{black}{\cite{decardi2018improving, harun2012dynamic}}:
\begin{equation}\label{reb:mass}
    \frac{d M_i}{d t} = \hat{F}_{in} m_{i,in} - \hat{F}_{V} q_{i,out} - \hat{F}_{L} m_{i, out}, 
\end{equation}
where $i = \text{CO}_2, \text{MEA}, \text{H}_2\text{O}$; $M_i$ represents the mass holdup of component $i$ in the reboiler in kmol; $\hat{F}_{in}$, $\hat{F}_{V}$, and $\hat{F}_{L}$ are the inlet stream, vapor, and liquid flow rates in kmol/s, respectively; $m_{i,in}$ and $m_{i,out}$ are the inlet and outlet liquid mole fractions of component $i$ in the reboiler; $q_{i,out}$ is the outlet vapor mole fraction of component $i$ in the reboiler.

\subsubsection{Energy balance}\label{reb:energy balance}

The dynamic behavior of the reboiler temperature $T_{reb}$ in Kelvin can be characterized as \textcolor{black}{\cite{decardi2018improving, harun2012dynamic}}:
\begin{equation}\label{reb:ene}
\begin{aligned}
    \rho_{reb} \hat{C}_{p,reb} V_{reb} \frac{d T_{reb}}{dt} &= \hat{F}_{in} H_{L,in} - \hat{F}_{V} H_{V,out} \\&~~~~- \hat{F}_{L} H_{L,out} + Q_{reb}
\end{aligned}
\end{equation}
where $\rho_{reb}$ represents the liquid density in the reboiler in kmol/m$^3$; $\hat{C}_{p,reb}$ represents the average heat capacity of the liquid in the reboiler in kJ/(kmol$\cdot$K); $V_{reb}$ is the holdup volume in the reboiler in m$^3$; $H_{L,in}$ and $H_{L,out}$ represent the liquid enthalpies at the inlet and outlet in kJ/kmol; $H_{V,out}$ is the enthalpy of vapor leaving the reboiler in kJ/kmol; $Q_{reb}$ denotes the reboiler heat duty in kW.

The reboiler heat duty combines the recovered heat energy and the heat from the diesel gas turbine, expressed as:
\begin{equation}\label{reb:Q}
    Q_{reb} = Q_{rec} + Q_{turbine}
\end{equation}

\begin{table}
\centering
\caption{Physical meanings of differential states of the shipboard PCC plant ($n=1,\ldots, 5$) \cite{zhang2025machine}}\vspace{2mm}
\label{table:diff states}\renewcommand\arraystretch{1.27}
\begin{tabular}{p{1.6cm}|p{1.6cm}|p{1.6cm}|p{1.6cm}}
\hline
State& Meaning & State & Meaning\\
\hline
\multicolumn{4}{c}{Absorption column}\\
\hline
{$x(1-5)$}&{$C_{L,\text{N}_2,A}^n$} &{$x(26-30)$}&  {$C^n_{G,\text{N}_2,A}$}\\
{$x(6-10)$}    &  {$C^n_{L,\text{CO}_2,A}$}          &{$x(31-35)$} &{$C^n_{G,\text{CO}_2,A}$}  \\
{$x(11-15)$}   &  {$C^n_{L,\text{MEA},A}$}       &{$x(36-40)$}& $C^n_{G,\text{MEA},A}$\\
{$x(16-20)$}  &  {$C^n_{L,\text{H}_2\text{O},A}$}  &{$x(41-45)$} &{$C^n_{G,\text{H}_2\text{O},A}$}\\
{$x(21-25)$}  & {$T_{L,A}^n$}               &{$x(46-50)$} & $T_{G,A}^n$\\
\hline
\multicolumn{4}{c}{Desorption column}\\
\hline
{$x(51-55)$}&{$C_{L,\text{N}_2, D}^n$}& {$x(76-80)$}&{$C^n_{G,\text{N}_2, D}$}\\
{$x(56-60)$}    &{$C^n_{L,\text{CO}_2, D}$}         &{$x(81-85)$}    &{$C^n_{G,\text{CO}_2, D}$}\\
{$x(61-65)$}    &{$C^n_{L,\text{MEA}, D}$}       &{$x(86-90)$}  &{$C^n_{G,\text{MEA}, D}$}\\
{$x(66-70)$}    &{$C^n_{L,\text{H}_2\text{O}, D}$}   &{$x(91-95)$}    &{$C^n_{G,\text{H}_2\text{O}, D}$}\\
{$x(71-75)$}  &{$T_{L, D}^n$}                          &{$x(96-100)$} &{$T_{G, D}^n$}\\
\hline
\multicolumn{4}{c}{Lean-rich solvent heat exchanger}\\
\hline
{$x(101)$} &{$T_{tube}$} &{$x(102)$} &{$T_{shell}$} \\
\hline
\multicolumn{4}{c}{Reboiler}\\
\hline
{$x(103)$}   &{$T_{reb}$}&&\\
\hline
\end{tabular}
\end{table}

\begin{table}[t]
  \renewcommand\arraystretch{1.2}
  \caption{Physical meanings of algebraic states of the shipboard PCC plant \cite{zhang2025machine}}\label{table:alg states}
  \centering
\begin{tabular}{p{1.6cm}|p{1.6cm}|p{1.6cm}|p{1.6cm}}
      \hline
      \multicolumn{4}{c}{Reboiler} \\ \hline 
      State &Meaning &State &Meaning \\ \hline
      $z(1)$ &$C_{L,\text{N}_2}$ &$z(5)$ &$q_{reb}$ \\
      $z(2)$ &$C_{L,\text{CO}_2}$ &$z(6)$ &$m_{\text{CO}_2, out}$ \\
      $z(3)$ &$C_{L,\text{MEA}}$ &$z(7)$ &$F_{G,reb}$ \\
      $z(4)$ &$C_{L,\text{H}_2\text{O}}$& \\ \hline     
    \end{tabular}
\end{table}

\subsection{Problem Formulation}
After discretization, the dynamic behaviors of the shipboard PCC system can be characterized by the following differential algebraic equations (DAEs) in discrete-time form \cite{zhang2025machine}:
\begin{subequations}\label{eq:process model}
\begin{align}
     x_{k+1}  &= f(x_{k},z_{k}, u_{k}, d_{k}) \label{eq:process model:f}\\
     0& = p(x_{k},z_{k}, u_{k}, d_{k})\label{eq:process model:g}\\
     y_{k} &= h(x_{k},d_{k})  \label{eq:process model:h}
\end{align}
\end{subequations}
where $x_k\in\mathbb{X}\subset\mathbb{R}^{103}$ is the state vector of the shipboard PCC system; $u_k\in \mathbb{U}\subset\mathbb{R}^{3}$ denotes the control inputs; $z_k\in\mathbb{Z}\subset\mathbb{R}^7$ denotes the algebraic states of the system; $y_k\in\mathbb{Y}\subset\mathbb{R}^4$ is the vector of controlled outputs; $d_k\in\mathbb{D}\subset\mathbb{R}^1$ denotes the known disturbance to the system. 
The physical interpretations of each element in the differential state vector and the algebraic state vector are given in Tables~\ref{table:diff states} and \ref{table:alg states} \cite{zhang2025machine}, respectively. Let $x(i)$ denote the $i$th element of state vector $x$. Among all the elements of $x$, only the temperatures of liquids and gas are measured in real-time, which are denoted as $\tilde{x}:= [x(21-25)^\text{T}, x(46-50)^\text{T}, x(71-75)^\text{T}, x(96-103)^\text{T}]^\text{T}$. The algebraic states are not measured in real-time. The concentration of $\text{CO}_2$ in the treated gas, denoted as $x({35})$, is measured to calculate the economic cost and the carbon capture rate of the process. We note that in the implementation of the proposed method, the measurement of this variable is only needed for offline model training and is not required during the implementation of the controller developed based on the Koopman model. 

The control input vector $u = [F_L, \tilde{F}_{fuel}, F_{sw}]^{\text{T}}$ includes the liquid solvent flow rate $F_L$ in m$^3$/s, the flow rate of fuel consumed by the diesel gas turbine, denoted by $\tilde{F}_{fuel}$, in kg/s, and the flow rate of seawater, denoted by $F_{sw}$, in m$^3$/s. The controlled output vector $y=[T_{reb}, p_{\text{CO}_2}, T^1_{L,A}, T^1_{L,D}]$ includes the reboiler temperature $T_{reb}$ in K, the carbon capture rate in percentage denoted by $p_{\text{CO}_2}$, the liquid temperature of the first layer of the absorption column, denoted by $T^1_{L,A}$, in K, and the liquid temperature of the first layer of the desorption column, denoted by $T^1_{L,D}$, in K. The carbon capture rate is calculated as follows:
\begin{subequations}
\begin{align}
p_{\text{CO}_2} &= \frac{\tilde{F}_{flue,\text{CO}_2} - \tilde{F}_{treated,\text{CO}_2}}{\tilde{F}_{flue,\text{CO}_2}}\\
\tilde{F}_{treated,\text{CO}_2} &= r_{\text{CO}_2} C^1_{G,\text{CO}_2,A} F_G \label{eq:co2release}
\end{align}
\end{subequations}
where \eqref{eq:co2release} characterizes the flow rate of CO$_2$ released into the atmosphere in kg/s. 

The economic operation of the shipboard PCC system aims to minimize the cumulative economic cost over a period of time, i.e.,
\begin{align}\label{empc:cost}
\sum_k&( \underbrace{\alpha \max(\tilde{F}_{treated,\text{CO}_2} - \bar{F}_{limit,\text{CO}_2},0)}_{c_k} + \beta u_{k}(2))
\end{align}
where $\alpha$ represents the carbon tax in $\$$/kg; $\beta$ is the price for the fuel in $\$$/kg; $\max(\tilde{F}_{treated,\text{CO}_2} - \bar{F}_{limit,\text{CO}_2},0)$ gives the quantity of CO$_2$ emitted beyond the CO$_2$ release threshold at time $k$ in kg/s; $\bar{F}_{limit,\text{CO}_2}$ is the CO$_2$ release threshold in kg/s, beyond which costs will be accrued; $u_k(2)$ represents the second variable of the control inputs, that is, the flow rate of fuel consumed by the diesel gas $\tilde{F}_{fuel}$. 

Necessary constraints are imposed to ensure operational safety and performance. Specifically, the reboiler temperature $T_{reb}$ is expected to be maintained between $385.15~\text{K}$ and $398.1~\text{K}$. $T^1_{L,A}$ is constrained to be lower than $353.15~\text{K}$ and $T^1_{L,D}$ is constrained to be larger than $353.15~\text{K}$, to increase the carbon capture rate of the system. 

In this work, we aim to develop a learning-based economic predictive control solution for the shipboard PCC system. By using an offline dataset $\mathcal{D}$ comprising data pairs $\left\{\tilde{x}_k, y_k, u_k, d_k, c_k\right\}$, the proposed approach learns an efficient predictive model for future key indices. Based on the learned model, an economic model predictive controller is developed for real-time economic operation of the PCC process. 

\begin{figure}
    \centering
    \includegraphics[width=0.38\textwidth]{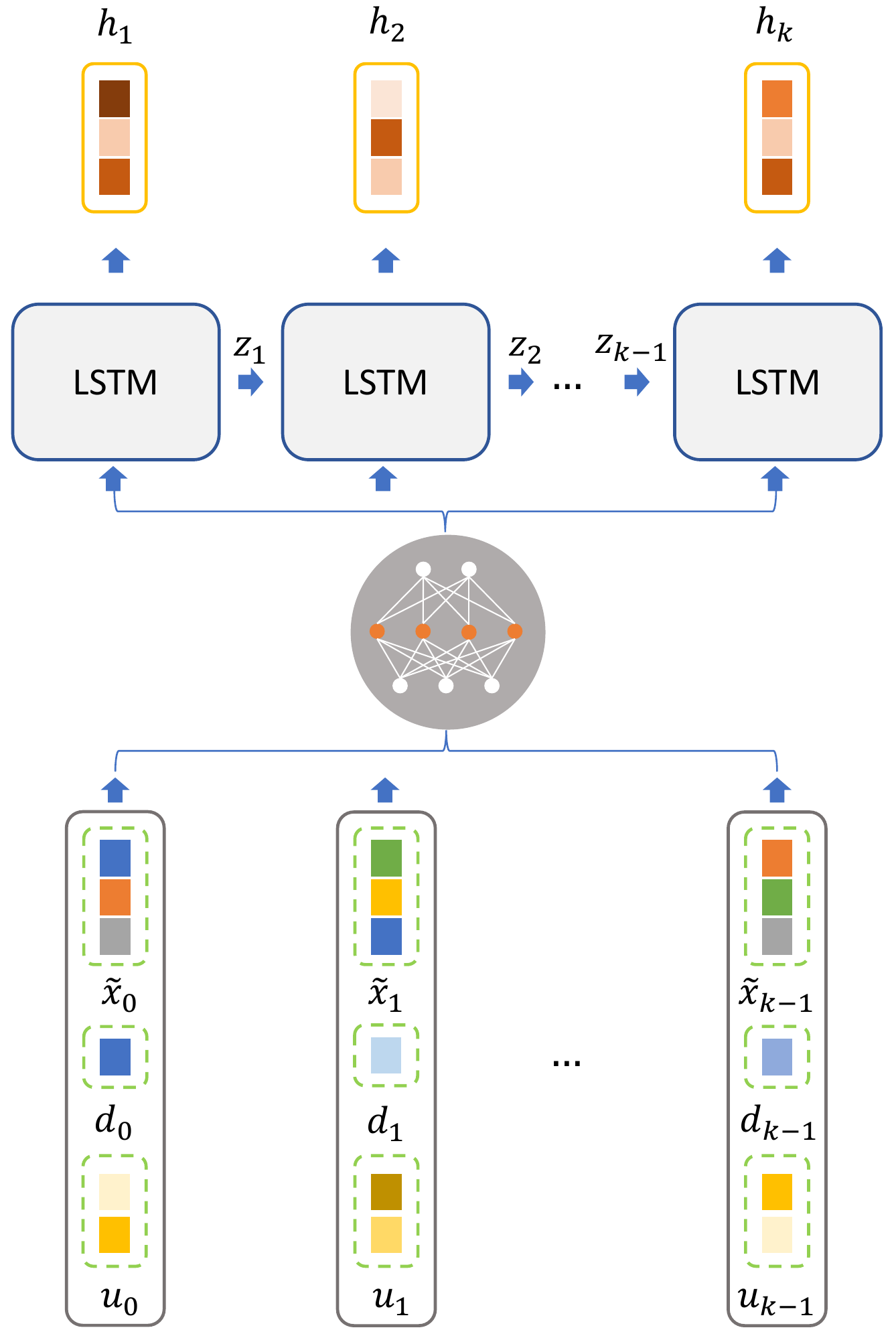}
    \caption{A schematic view of the LSTM-based observable function. }
    \label{fig:RNN}
\end{figure}
\begin{figure*}
    \centering
    \includegraphics[width=0.83\textwidth]{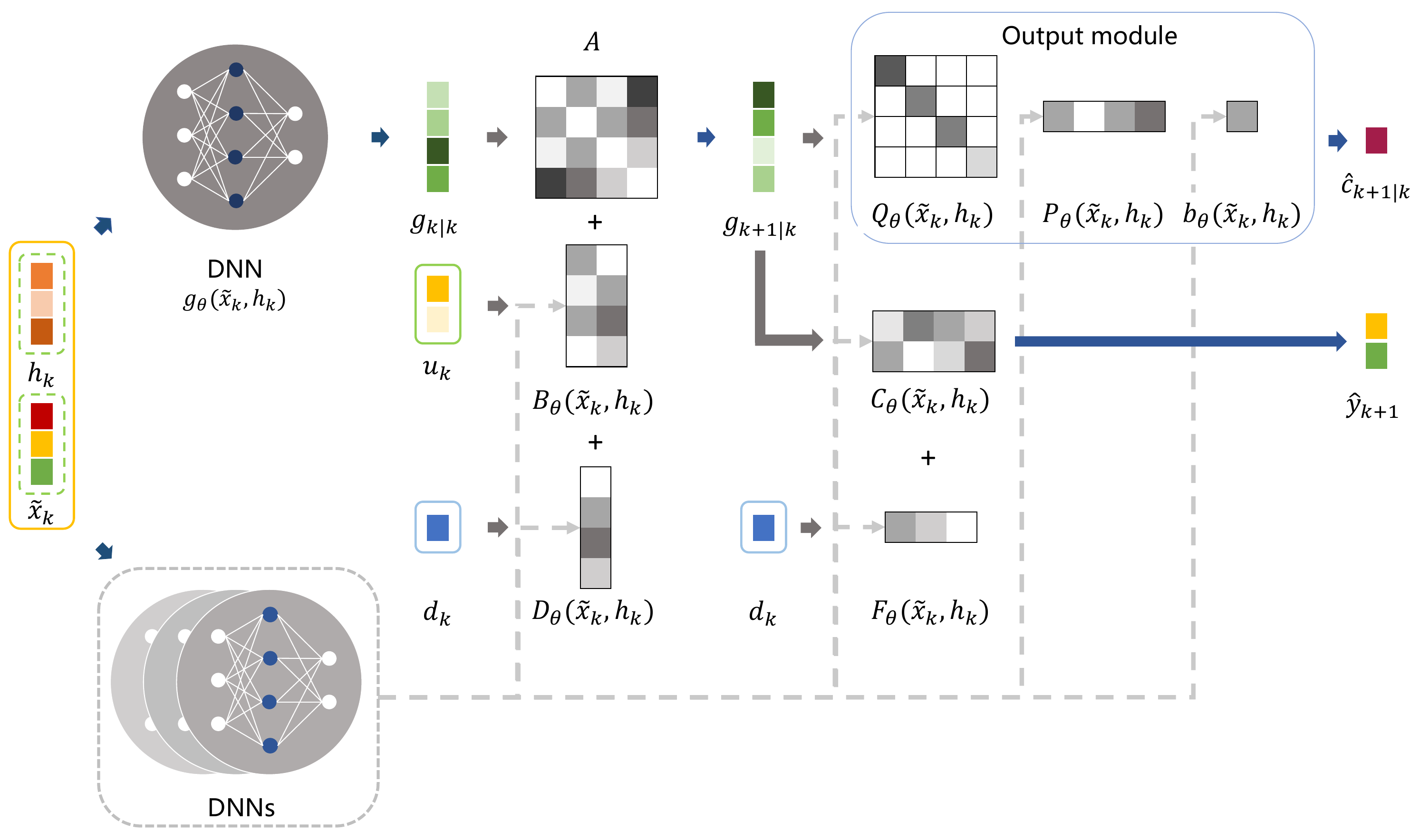}
    \caption{An overview of the proposed DNKO model structure. }
    \label{fig:pipeline}
\end{figure*}
\section{Deep Neural Koopman Modeling}\label{sec:DNKO}
In this section, we introduce the details of the building blocks of the deep neural Koopman operator (DNKO) model and the training of the time-varying model. 

\subsection{The Koopman operator theory}
In this subsection, we briefly introduce the ideas and notations of the Koopman operator theory. The Koopman theory was first formulated in \cite{koopman1931hamiltonian}. 
According to the Koopman theory, a general nonlinear system of the following form: 
\begin{equation*}
    x_{k+1} = f(x_k), k\in \mathbb{N}
\end{equation*}
can be transformed into a linear system within an infinite-dimensional function space $\mathcal{G}$. 
This space encompasses all square-integrable real-valued functions defined over the compact domain $\mathbb{X}\subseteq\mathbb{R}^{n}$. The elements of $\mathcal{G}$, denoted as $g$, are referred to as \textit{observables}. The Koopman operator $\mathcal{K}:\mathcal{G} \rightarrow \mathcal{G}$ satisfies the following relationship:
\begin{equation*}
g\circ f(x_k) = \mathcal{K}g(x_k)
\end{equation*}
where $\circ$ denotes function composition, and $g\in \mathcal{G}$ represents the observable function. The use of the Koopman operator enables the prediction of future states of the nonlinear system in a linear fashion as follows:
\begin{equation*}
x_{k+1} = g^{-1}(\mathcal{K}g(x_k))
\end{equation*}
In practical applications, it is often more manageable to work within a finite-dimensional function space $\overline{\mathcal{G}}\subset \mathcal{G}$. This space is defined by a set of linearly independent observables $\{g|g:\mathbb{R}^{n}\rightarrow\mathbb{R}^h\}$, resulting in a finite-dimensional Koopman operator $K$. While initially proposed for autonomous nonlinear systems, the concept of the Koopman operator has been extended to controlled systems in recent years \cite{korda2018linear,proctor2018generalizing}.
For controlled systems, the Koopman operator adheres to the following condition:
\begin{equation}\label{eq:classical controlled Koopman}
g_x\circ f(x_k, u_k) = A g_x(x_k) + B g_u(x_k, u_k),
\end{equation}
where $A\in \mathbb{R}^{h\times h}$ and $B\in \mathbb{R}^{h\times r}$ are submatrices of the Koopman operator, and $g_x$ and $g_u$ represent the observables for the state $x_k$ and control input $u_k$, respectively.

\subsection{Observable Encoding}

In Koopman modeling, the encoding of observables plays a crucial role in the predictive capability of the resulting model. For the shipboard PCC process characterized by the set of DAEs in \eqref{eq:process model}, the future states are not only dependent on the current state and inputs, but they are also dependent on the algebraic states of the process hidden from the dataset. To overcome the obstacle of lacking algebraic states data, we adopt Long Short-Term Memory (LSTM) \cite{hochreiter1997long}, to infer the information contained in the algebraic states from the trajectory of previous states and inputs.

As shown in Fig.~\ref{fig:RNN},  the state and input vectors are concatenated and fed into a pre-encoding deep neural network (DNN) \cite{lecun2015deep}, which integrates the states and inputs to extract useful information for future prediction. The output vector of the DNN is recurrently fed into the LSTM to form the encoded hidden state $h\in \mathbb{R}^l$. 
At time step $k$, the hidden state $h_k$ is produced by feeding $\tilde{x}_{k-\tau:k-1}$, $d_{k-\tau:k-1}$ and $u_{k-\tau:k-1}$ over the previous $\tau$ steps into the pre-encoding DNN and LSTM. $\tau$ is a tunable hyperparameter to trade-off modeling performance and computational complexity. Finally, the observables $g_{k|k}$ are obtained by feeding $\tilde{x}_k$ and $h_k$ into the post-encoding DNN $g_\theta(\cdot)$, that is, $g_{k|k} = g_\theta(\tilde{x}_k, h_k)$. Here and thereafter, we use $\theta$ to denote the trainable parameters of the DNNs mentioned in the proposed pipeline.

\subsection{Prediction}\label{subsec:Koopman-predition}

In this subsection, we show how to exploit the encoded observables to predict the future stage cost $c$ and control outputs $y$. As shown in \eqref{eq:classical controlled Koopman}, the observables propagate linearly with the Koopman operators. In this work, we propose to use the following time-varying Koopman model structure to predict future observables:
\begin{equation}\label{eq:time-varying Koopman}
    g_{k+j+1|k} = A g_{k+j|k} + B_ku_{k+j} + D_kd_{k+j}
\end{equation}
where $j\in\mathbb{Z}_{[0, T_f]}$ in which $\mathbb{Z}_{[0, T_f]}$ represents the set of integers between 0 and $T_f$, with $T_f$ denoting the prediction horizon; $B_k = B_\theta(\tilde{x}_k,h_k)\in\mathbb{R}^{h\times 3}$ and $D_k = D_\theta(\tilde{x}_k,h_k)\in\mathbb{R}^{h\times 1}$ are time-varying matrices generated by the DNNs \cite{lecun2015deep} in Fig.~\ref{fig:pipeline}, with inputs to the DNNs being $\tilde{x}_k$ and $h_k$, respectively. We note that for each time instant $k$, $B_k$ and $D_k$ are generated by their respective DNNs once, and these two matrices remain fixed within the $T_f$-step prediction horizon for time instant $k$. 

From the predicted observables, the future stage costs and controlled outputs are reconstructed as follows:
\begin{subequations}\label{eq:reconstruction}
    \begin{align}
        \hat{c}_{k+j|k}  &= g_{k+j|k}^\text{T} Q_k g_{k+j|k} + P^\text{T}_kg_{k+j|k} + b_k,\label{eq:reconstruction-cost}\\
        \hat{y}_{k+j|k} & = C_k g_{k+j|k} + F_k d_{k+j},\label{eq:reconstruction-output}
    \end{align}
\end{subequations}
where $j\in\mathbb{Z}_{[0, T_f]}$.
In \eqref{eq:reconstruction}, $Q_k = \text{diag}\{Q_\theta(\tilde{x}_k,h_k)\}\in\mathbb{R}^{h\times h}$ is a diagonal weighting matrix encoded by $Q_\theta(\cdot)$, of which the output layer has ReLU as the activation function, such that $Q_k$ is guaranteed to be semi-positive-definite. $\text{diag}\{\cdot\}$ signifies a diagonal matrix with the given vector as its diagonal elements. $P_k=P_\theta(\tilde{x}_k, h_k)\in\mathbb{R}^{h}$ is an encoded weighting vector, and $b_k = b_\theta(\tilde{x}_k, h_k)\in\mathbb{R}$ represents an encoded scalar bias. In \eqref{eq:reconstruction-output}, the output matrix  $C_k=C_\theta(\tilde{x}_k, h_k)$ and disturbance matrix $F_k=F_\theta(\tilde{x}_k, h_k)$ are also encoded by the DNNs, respectively. Similar to \eqref{eq:time-varying Koopman}, the encoded matrices here remain fixed within the prediction horizon for each time instant $k$. An illustration of the structure of the proposed DNKO model is given in Fig.~\ref{fig:pipeline}.

The Koopman operator model presented in \eqref{eq:time-varying Koopman} and \eqref{eq:reconstruction} preserves the linearity of observable propagation and output reconstruction. Furthermore, the predicted stage costs in \eqref{eq:reconstruction-cost} are convex with respect to the observables and the control inputs. These features all together ensure the convexity of the resulting predictive control problem, which will be introduced in the next section.

\begin{rmk}
{\color{black}In this work, the reconstruction of future stage costs in \eqref{eq:reconstruction-cost} is inspired by that in \cite{han2024efficient} where a quadratic function with constant weighting matrices and parameters is constructed in a lifted space to approximate the nonlinear economic state cost for a water treatment plant (see Section 4 of \cite{han2024efficient}).
As is different from \cite{han2024efficient}, the weighting matrices and parameters in \eqref{eq:reconstruction-cost}, including $Q_k$, $P_k$ and $b_k$, as well as the Koopman matrices $B_k$ and $D_k$ are encoded by the respective neural networks, and they vary over time. At each time step, these parameters are updated. This way, the modeling performance can be improved compared to the case when constant Koopman matrices and parameters are used \cite{han2024efficient}. 
Additionally, as compared to \cite{han2024efficient}, the current method reconstructs the controlled outputs as described in (\ref{eq:reconstruction-output}); this will enable the Koopman-based controller to handle hard constraints on the system outputs, which will be discussed in the next section.
We note that the approach proposed in this work also preserves the convexity of the resulting control problem, which facilitates efficient real-time control. }
\end{rmk}

\subsection{Training}

In the DNKO model, $A$ and the parameters of the DNNs $\theta$ are trained by minimizing the following finite-horizon objective function:
{
\begin{subequations}\label{eq:loss}
\begin{align}
        \min_{\theta, A} \text{    }& \mathbb{E}_\mathcal{D} \sum_{j=1}^{T_f} \left(\Vert y_{k+j}- \hat{y}_{k+j|k} \Vert_2^2 +\Vert c_{k+j}-\
        \hat{c}_{k+j|k}\Vert_2^2 \right) \label{eq:obj1}\\ 
        \text{s.t. }& g_{k+j+1|k} = A g_{k+j|k} + B_ku_{k+j} + D_kd_{k+j}, \\
        &\hat{c}_{k+j|k}  = g_{k+j|k}^\text{T} Q_k g_{k+j|k} + P^\text{T}_kg_{k+j|k} + b_k,\\
        &\hat{y}_{k+j|k} = C_k g_{k+j|k} + F_k d_{k+j},
        \\
        &g_{k|k} = g_\theta(x_{k}, h_{{k}})
\end{align}
\end{subequations}}

\noindent
where 
$\mathcal{D}:= \{[\tilde{x}_k, u_k, d_k, y_k, c_{k}]^i_{0:T_f+\tau}\}_{i=1:N}$ denotes the sampled dataset; $\mathbb{E}_\mathcal{D}(\cdot)$ returns the expectation of variables over the data distribution of $\mathcal{D}$. The matrices $B_k$, $C_k$, $D_k$, $F_k$, $Q_k$, $P_k$, and scalar $b_k$, are encoded from $\tilde{x}_k$ and $h_k$ using the corresponding neural networks described in the Section~\ref{subsec:Koopman-predition}.
In objective function \eqref{eq:obj1}, the first term and the second term correspond to the prediction errors for the system output and economic cost, respectively.
The optimization problem in \eqref{eq:loss} is solved using Adam \cite{kingma_adam_2017}, which is a stochastic gradient descent algorithm. \textcolor{black}{At each iteration, a batch of data is sampled from the dataset, and stochastic gradient descent is conducted over the sampled batch of data. To evaluate generalization and convergence, the model performance is periodically evaluated on both a validation dataset and a test dataset after a predefined number of iterations. Additionally, the learning rate $\delta$ is decayed by a discount factor $\gamma$ at regular intervals, determined by the discount steps $N_d$, to facilitate convergence. The training process is summarized in the pseudo-code in Algorithm~\ref{algo:DNKO}. }

{\color{black}
\begin{algorithm}[h]
   \caption{Training of Deep Neural Koopman Operator}
   \label{algo:DNKO}
{\color{black}
\begin{algorithmic}
    \REQUIRE
    Dataset $\mathcal{D}$, prediction horizon $T_f$, history horizon $\tau$, learning rate $\delta$, discount factor $\gamma$, the discount steps $N_d$, batch size, the maximum episode number $N$, and the steps per epoch $M$
   \FOR{$i=1$ to $N$}
   \FOR{$j=1$ to $M$}
   \STATE Sample a batch of data from $\mathcal{D}$
   \STATE Update $\theta, A$ by using stochastic gradient descent with respect to the objective function \eqref{eq:loss}
   \ENDFOR
   \STATE Evaluate the modeling performance
   \IF{there exists $l\in \mathbb{N}$ such that $l*i = N_d$}
   \STATE $\delta \leftarrow \gamma * \delta$
   \ENDIF
   \ENDFOR
\end{algorithmic}
}
\end{algorithm}
}

\section{Economic Model Predictive Control}\label{sec:empc}

Based on the established DNKO model, we develop a data-driven constrained EMPC method for the economic operation of the shipboard PCC system. The optimization problem associated with the EMPC design is presented as follows:

\begin{subequations}\label{eq:economic MPC}
\begin{align}
    \min_{u_{k:k+T_f-1|k}} \sum_{j=1}^{T_f} \big(& \hat{c}_{k+j|k} + \beta u_{k+j}(2)\big) + \sum_{j=1}^{T_f-1}\left\Vert\Delta u_{k+j}\right\Vert_R\\
    \text{s.t. } ~g_{k+j+1|k} &= A g_{k+j|k} + B_ku_{k+j} + D_kd_{k+j} \label{eq:economic MPC1}\\
    \hat{c}_{k+j|k}  &= g_{k+j|k}^\text{T} Q_k g_{k+j|k} + P^\text{T}_kg_{k+j|k} + b_k\\
    \hat{y}_{k+j|k} & = C_k g_{k+j|k} + F_k d_{k+j}\\
    \Delta u_{j|k} &= u_{j+1|k} - u_{j|k}\\
    g_{k|k} &= g_\theta(\tilde{x}_k, h_{{k}})\label{eq:economic MPC2}\\
    u_{k+j}&\in \mathbb{U}\label{eq:input constraint}\\
    y_{k+j|k}&\in \mathbb{Y}\label{eq:output constraint}\\
    \bar{p}_{\text{CO}_2} &\leq \sum_{j=1}^{T_f}y_{k+j|k}(2)/T_f \label{eq:capture rate constraint}
\end{align}
\end{subequations}
The second term in (\ref{eq:economic MPC1}), where $R$ is a positive definite matrix, penalizes the rate of change in the control inputs. The optimization problem \eqref{eq:economic MPC} is solved to obtain the optimal control input sequence $u^*_{k:k+T_f-1|k}$. At each sampling instant, only the first control action $u^*_{k|k}$ of the optimal control sequence is applied to the underlying nonlinear system in (\ref{eq:process model}). \textcolor{black}{In (\ref{eq:capture rate constraint}), $y(2)$ represents the second variable of the controlled output vector, i.e., the carbon capture rate in percentage.} 

In \eqref{eq:economic MPC}, Koopman matrices $B_k$ and $D_k$, and initial observable vector are encoded from DNNs using the current partial state $\tilde{x}_k$ and the $h_{k}$ generated by the LSTM. \eqref{eq:input constraint} and \eqref{eq:output constraint} are inequality constraints imposed on the control inputs and controlled outputs, respectively. Furthermore, a constraint on the average carbon capture rate is introduced in \eqref{eq:capture rate constraint} to maintain the overall carbon capture rate above the pre-specified lower bound $\bar{p}_{\text{CO}_2}$. Meanwhile, it is worth mentioning that the carbon capture rate may fluctuate significantly due to variations in the engine loads, which may render the optimization problem in (\ref{eq:economic MPC}) infeasible due to \eqref{eq:capture rate constraint}. If infeasibility is encountered during solving \eqref{eq:economic MPC}, we may alternatively solve a soft constrained optimization problem which is formulated by including a slack variable in \eqref{eq:capture rate constraint}, shown in the following form:
\begin{subequations}\label{eq:slacked economic MPC}
\begin{align}
    \min_{u_{k:k+T_f-1|k}, s} &\sum_{j=1}^{T_f} \big( \hat{c}_{k+j|k} 
    + \beta u_{k+j}(2)\big) \notag\\&+ \sum_{j=1}^{T_f-1}\left\Vert\Delta u_{k+j}\right\Vert_R + w_s s^2\\
    \text{s.t. } ~&\eqref{eq:economic MPC1}-\eqref{eq:output constraint}\\
    &\bar{p}_{\text{CO}_2} - s \leq \sum_{j=1}^{T_f}y_{k+j|k}(2)/T_f \\
    &s\geq 0
\end{align}
\end{subequations}
where $s\in\mathbb{R}^+$ is a positive slack variable to be optimized, and $w_s$ is a pre-determined weight penalizing the violation of constraints. 


\begin{rmk}
{\color{black}    
The computational complexity of the DNKO-based EMPC primarily arises from two key components: evaluating the neural networks to encode the time-varying Koopman matrices $B_k$ and $D_k$ in (\ref{eq:time-varying Koopman}) and (\ref{eq:economic MPC1}), and solving the optimization problem associated with the EMPC. Specifically, the neural networks are used to parameterize the observable encoding and time-varying matrices. The complexity of the neural network evaluation depends on the numbers of layers and neurons of these neural networks. The complexity of solving the optimization problem associated with the proposed EMPC is affected by factors such as the dimension of the observable vector, the length of the prediction horizon, and the number of constraints.}
The proposed method holds the potential for achieving economic operations of high-dimensional nonlinear systems/processes in a computationally efficient manner. 
\end{rmk}

\begin{table}[tb]
  \renewcommand\arraystretch{1.19}
  \caption{System parameters of shipboard carbon capture plant \cite{wartsila46f,ben2023desulfurization}}\label{table:pcc:config}
  \centering
    \begin{tabular}{ l l l }
      \toprule
      \bfseries Parameter                            & \bfseries Value      \\ \midrule
      \bfseries Absorption and desorption column         \\
      Internal diameter of absorption column $D_c$  & 4.2~m \\
      Internal diameter of desorption column $D_c$  & 4.9~m \\
      Height of absorption and desorption columns                  & 12.5~m \\
      Column interfacial area $a^I$   & 143.9~m$^2$/m$^3$ \\ 
      Nominal packing size           & 0.038~m \\ \hline
      \bfseries Lean-rich solvent heat exchanger                  \\
      Volume of tube side $V_{tube}$   & 0.0155~m$^3$ \\
      Volume of shell side $V_{shell}$  & 0.4172~m$^3$ \\
      Overall heat transfer coefficient $U$  &1899.949~kW/K \\  \hline
      \bfseries Seawater heat exchanger       \\
      Inlet seawater temperature $T_{sw,in}$  &   308~K   \\
      Outlet seawater temperature $T_{sw,out}$  &     323~K   \\ \hline
      \bfseries Reboiler \\
      Holdup volume in reboiler $V_{reb}$    & 0.145~m$^3$ \\ \hline
      \bfseries Ship engines \\
      Engine output $Q_E$ with 100$\%$ engine load  & 10800~kW \\  \hline
      \bfseries Fuel component \\
      Mole fraction of carbon in the fuel $q_{fuel,C}$ & $84.86\%$ \\ \hline
      \bfseries Flue gas component \\
      Mole fraction of CO$_2$ in the flue gas $q_{flue,\text{CO}_2}$ & $5.462\%$ \\  \hline
    \bfseries  \textcolor{black}{Tax and price}\\
      \textcolor{black}{Carbon tax $\alpha$} & \textcolor{black}{0.05 $\$$/kg} \\ 
      \textcolor{black}{Price for the fuel $\beta$} & \textcolor{black}{1.2852 $\$$/kg}\\
       \bottomrule
    \end{tabular}
\end{table}

\begin{table}[tb]
  \renewcommand\arraystretch{1.25}
  \caption{The upper bounds and lower bounds of control inputs $u$ and known disturbance $d$.}\label{table:up:bound}
  \centering
    \begin{tabular}{ c c c c c }
      \toprule
             &{$F_L$ (m$^3$/h)} &{$\tilde{F}_{fuel}$ (kg/h)} &{$F_{sw}$ (m$^3$/h)} &{$\varphi_E$ $(\%)$} \\ \midrule
      Upper bounds &$120$        &$5000$                    &$90$           &$100$\\
      Lower bounds &$30$        &$2500$                     &$20$            &$20$\\ \bottomrule
    \end{tabular}
\end{table}

\section{Results}\label{sec:results}

We apply the proposed DNKO-based EMPC method to the simulated shipboard PCC system \cite{zhang2025machine}. The model accuracy and control performance of the proposed method are evaluated. Additionally, we compare the control performance of the proposed method with two baselines: proportional-integral (PI) control designed to track the economically optimal steady-state, and the state-of-the-art model-free reinforcement learning method, soft actor-critic (SAC) \cite{haarnoja2018soft}.

\begin{figure}[htb]
    \centering
\includegraphics[width=0.46\textwidth]{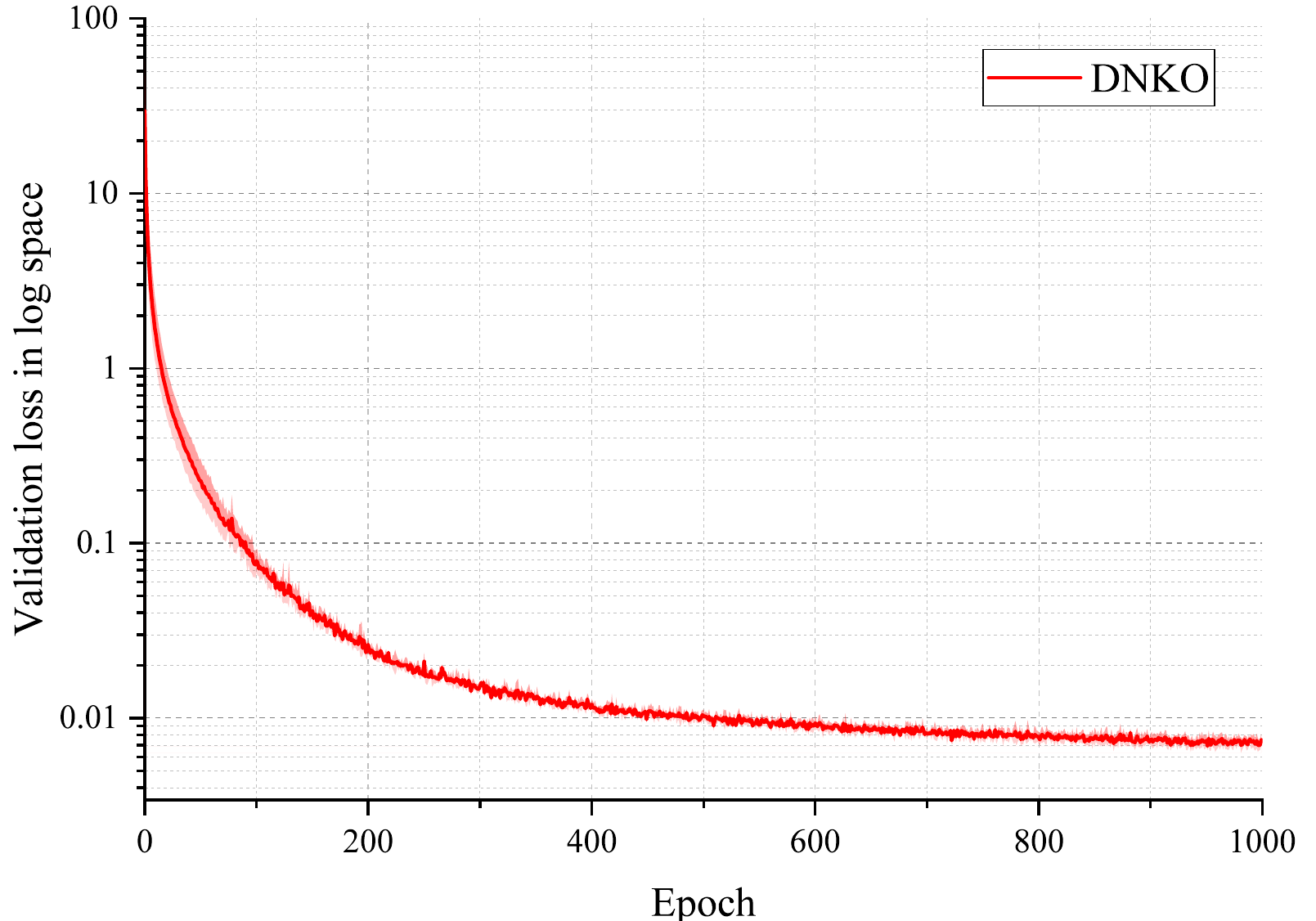}
\centering
\includegraphics[width=0.46\textwidth]{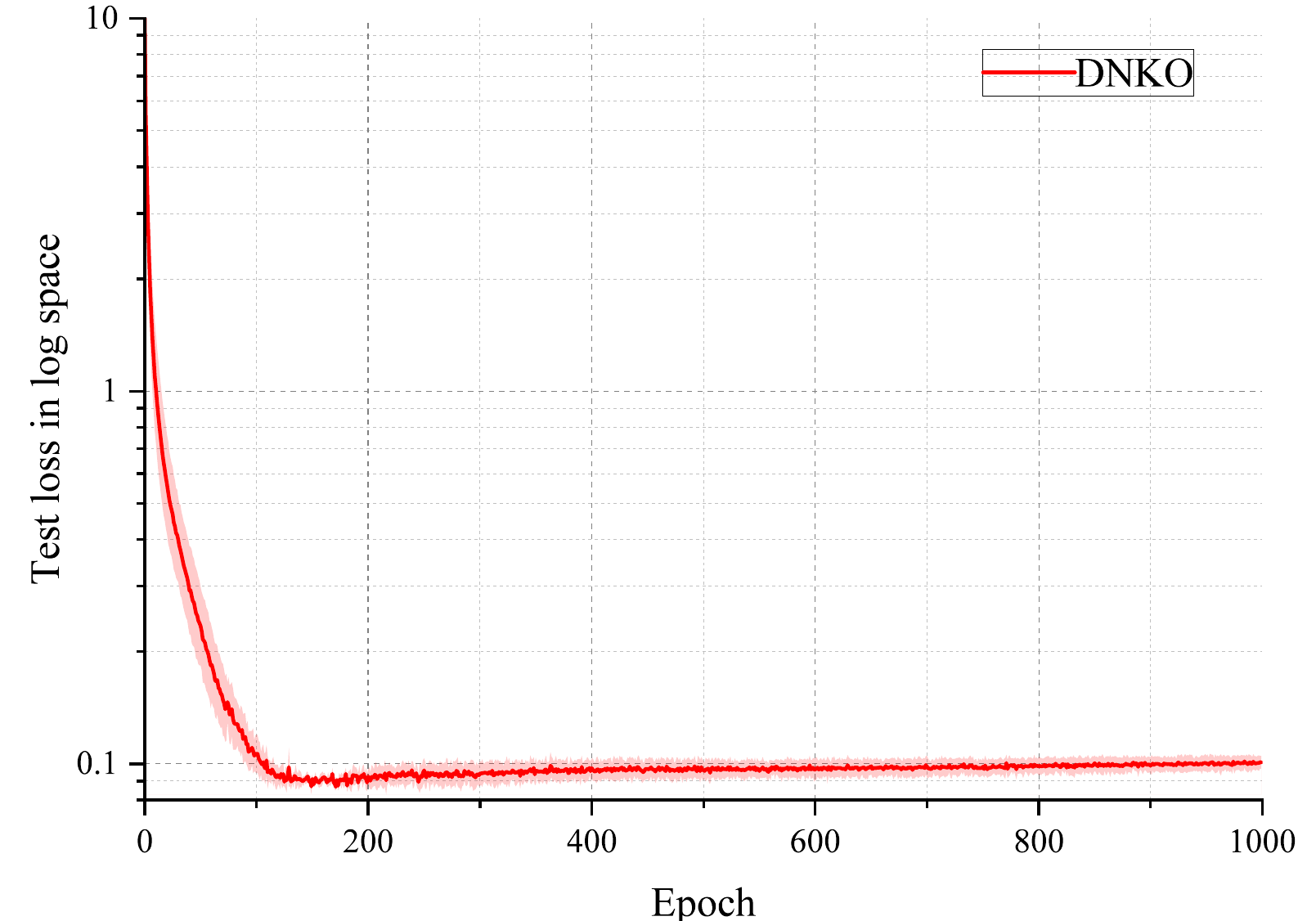}
    \caption{Cumulative prediction errors given by the DNKO modeling approach on the validation and test datasets. The X-axis represents the number of training epochs; the Y-axis shows the cumulative mean-squared prediction error on a logarithmic scale over 16 time steps. The shaded region depicts the confidence interval, which is generated based on one standard deviation calculated based on 10 random initializations.}
    \label{fig:modeling performance}
\end{figure}

\begin{table}[t]
  \renewcommand\arraystretch{1.25}
  \caption{\textcolor{black}{Four} ship operational conditions and engine load ranges under each condition (adapted from \cite{faber2012regulated,chen2021operational,pelic2023impact}).}\label{table:engine load: exp}
  \centering
    \begin{tabular}{p{3.5cm} p{3.5cm}}
      \toprule
      Operational condition     & Range of engine load $\varphi_E$ $(\%)$ \\ \midrule
      Maneuvering    &70$\%$ - 100$\%$ \\
      Slow steaming    &40$\%$ - 70$\%$\\
      Low engine load    & 20$\%$ - 40$\%$ \\ 
      \textcolor{black}{Full range operation}    &\textcolor{black}{20$\%$ - 100$\%$} \\ 
      \bottomrule 
    \end{tabular}
\end{table}

\begin{table}[t]
  \renewcommand\arraystretch{1.25}
  \setlength{\tabcolsep}{18pt}
\caption{\textcolor{black}{Hyperparameters of DNKO}}\label{table:hyperparameters}
\centering
\begin{tabular}{l l}
\toprule
\textcolor{black}{Hyperparameters}&\textcolor{black}{Value}\\
\midrule
\textcolor{black}{Batch Size} & \textcolor{black}{$128$}\\
\textcolor{black}{Learning rate} & \textcolor{black}{$5\times 10^{-3}$}\\
\textcolor{black}{Prediction horizon $H$} & \textcolor{black}{$16$} \\
\textcolor{black}{Structure of DNNs} & \textcolor{black}{$(128,128)$}\\
\textcolor{black}{Number of LSTM cells} &\textcolor{black}{$16$}\\
\textcolor{black}{History horizon $\tau$} &\textcolor{black}{$4$} \\
\textcolor{black}{Activation function} & \textcolor{black}{ReLU}\\
\textcolor{black}{Dimension of observables} & \textcolor{black}{$40$}\\
\textcolor{black}{$l_2$ norm regularization coefficient} & \textcolor{black}{$0.1$}\\
\textcolor{black}{Penalization weight $w_s$ } & \textcolor{black}{$10^6$}\\
\bottomrule
\end{tabular}
\end{table}

\subsection{Simulation settings}

\begin{figure}
    \subfloat[Condition 1 -- Maneuvering]{\includegraphics[width=0.46\textwidth]{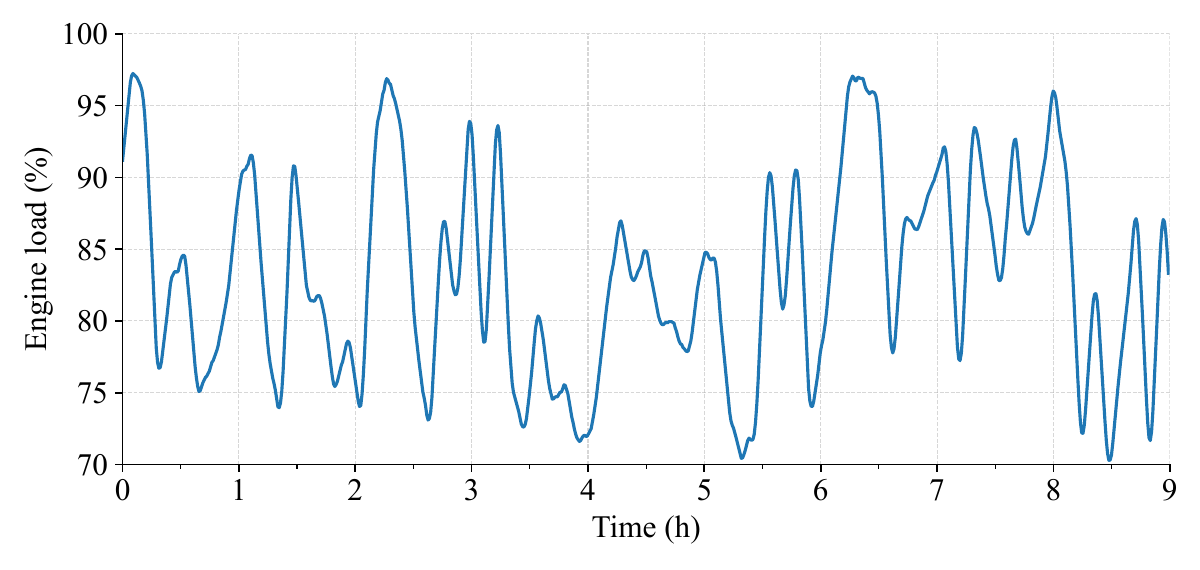}}\\
    \subfloat[Condition 2 -- Slow steaming]{\includegraphics[width=0.46\textwidth]{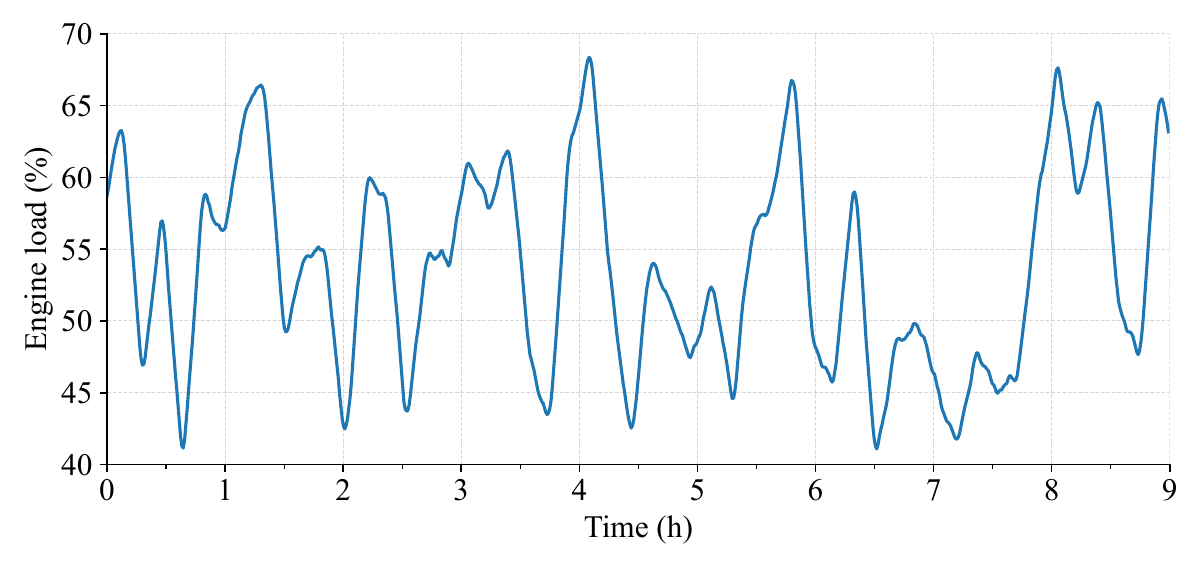}}\\
    \subfloat[Condition 3 -- Low engine load]{\includegraphics[width=0.46\textwidth]{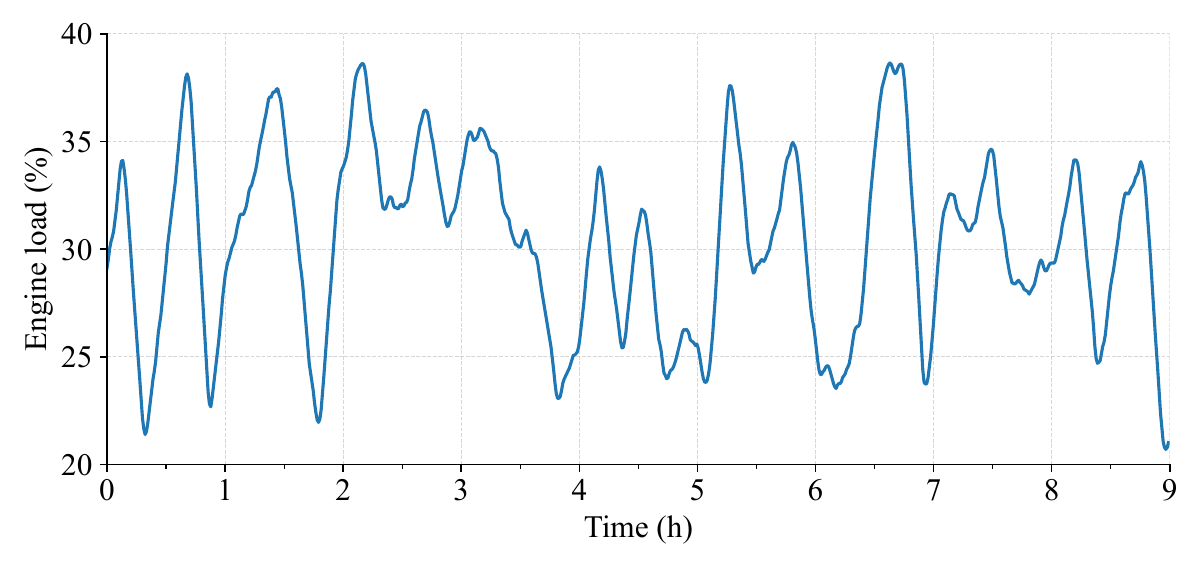}}\\
    \subfloat[\textcolor{black}{Condition 4 -- Full range opeartion} ]{\includegraphics[width=0.46\textwidth]{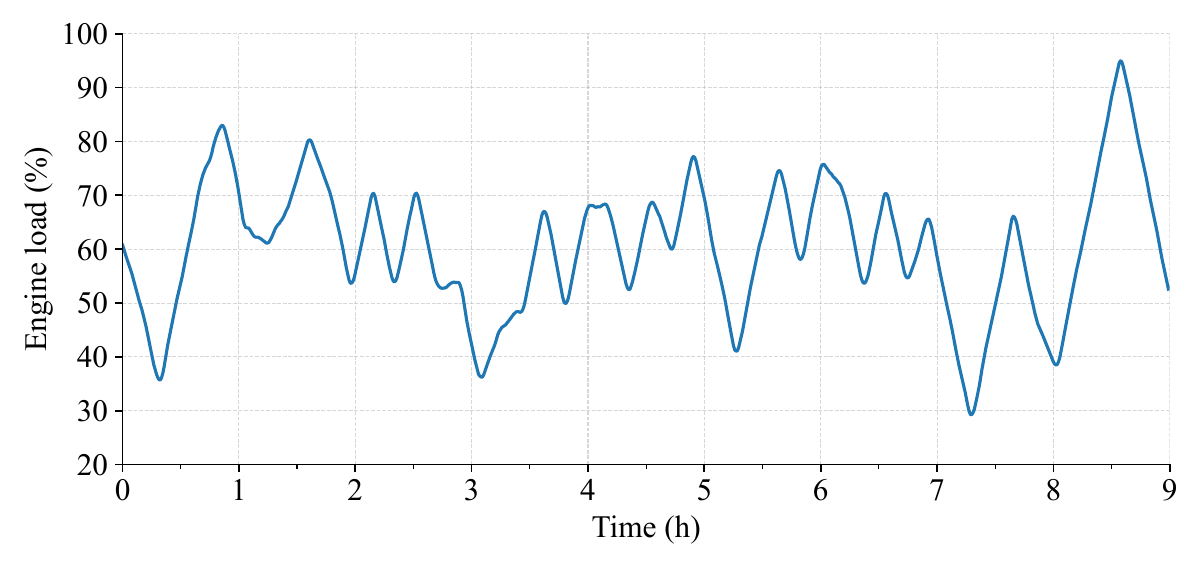}}
    \caption{Engine load trajectories under the \textcolor{black}{four} operational conditions.}
    \label{fig:engine_load}
    \vspace{-2em}
\end{figure}

Table \ref{table:pcc:config} presents the key parameters for the shipboard post-combustion carbon capture system, which uses IMTP \#40 packing. The values of the parameters for the ship engines are sourced from \cite{wartsila46f}, while the fuel component, $n$-hexadecane, is referenced in \cite{ben2023desulfurization}. We generate the flue gas components, which are given in in Table \ref{table:pcc:config}, from simulations using Aspen Plus at different ship engine load levels. The first-principles dynamic model described in Section~\ref{sec:system}, along with the model parameters in Table~\ref{table:pcc:config}, serves as a high-fidelity dynamic process simulator for data generation. The process is simulated using an implicit differential-algebraic solver (IDAS) integrator, which is designed for solving DAE systems, within the CasADi SUNDIALS suite \cite{andersson2019casadi}.

The process simulation is conducted with a sampling period of $40$ s. The engine load varies within the range of $20\%$ to $100\%$. The ranges of control inputs and engine loads are detailed in Table~\ref{table:up:bound}. \textcolor{black}{During data collection, control inputs are sampled from the input space at random intervals. Each sampled input is held constant for a random duration  (ranging from 10 to 30 steps); this is to achieve temporal diversity in the dataset. Gaussian noise is added to the control signals to enhance data richness. These inputs are constrained to remain within the input space. Engine load trajectories are generated as piecewise linear paths between randomly sampled target values. Gaussian noise is added to the input trajectories, and the values of the inputs are constrained to remain within $20\% - 100\%$.} In total, a dataset comprising $2 \times 10^4$ samples is collected, of which $90\%$ are used for model training, while the remaining $10\%$ are allocated for validation. Additionally, a test dataset containing $800$ data samples is collected. All the data samples are normalized using the mean and standard deviation vectors of the collected data. In the evaluation of control performance, \textcolor{black}{four} representative ship operational conditions \textcolor{black}{adapted (with slight changes) from \cite{faber2012regulated,chen2021operational,pelic2023impact}} are considered: maneuvering (Operational Condition 1), slow steaming (Operational Condition 2), low engine load (Operational Condition 3), \textcolor{black}{and full range operation (Operational Condition 4)}. The definitions of the \textcolor{black}{four} different operational conditions, and the engine load variation ranges under these conditions are presented in Table~\ref{table:engine load: exp}.

\begin{figure}
    \hspace{1.4cm}\includegraphics[width=0.35\textwidth]{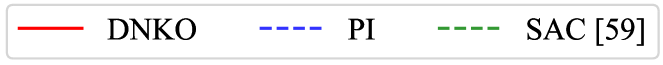}
    \subfloat[Condition 1 -- Maneuvering]{\includegraphics[width=0.47\textwidth]{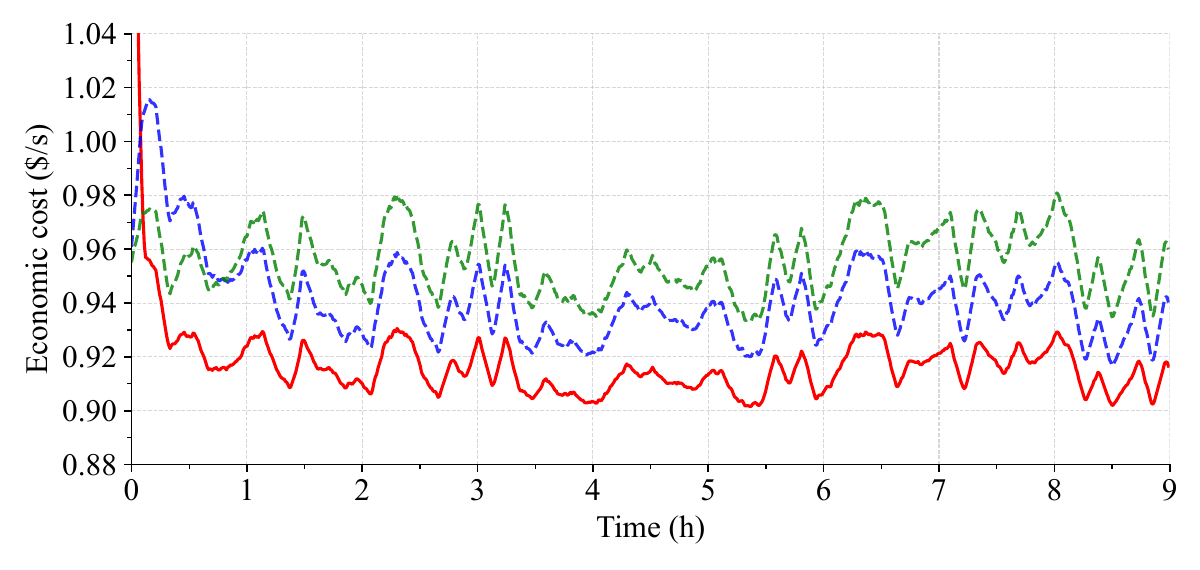}}\\
    \subfloat[Condition 2 -- Slow steaming]{\includegraphics[width=0.47\textwidth]{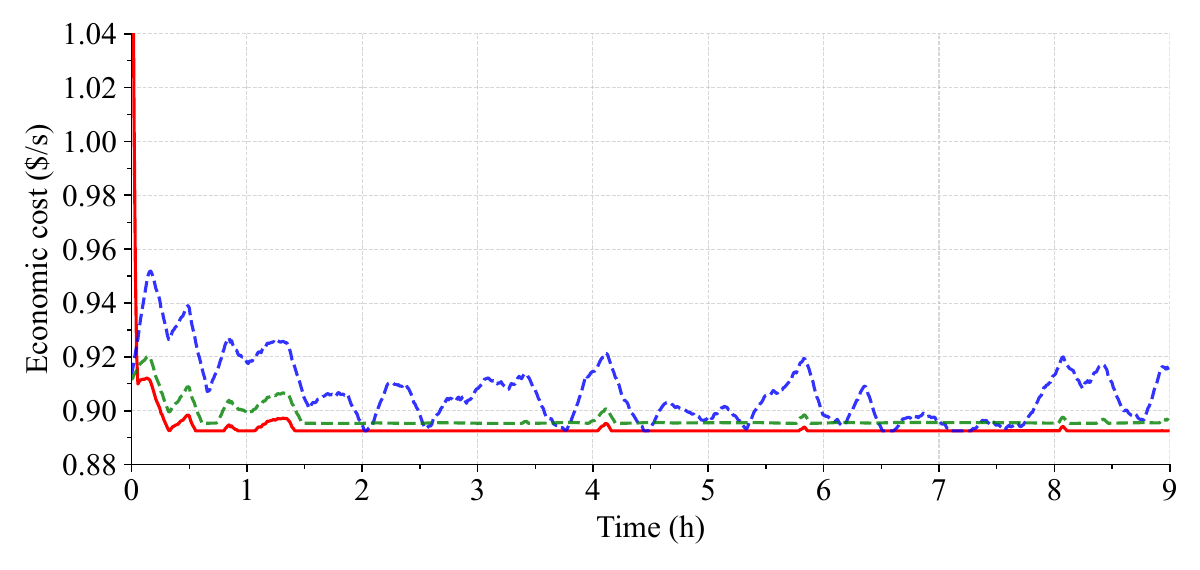}}\\
    \subfloat[Condition 3 -- Low engine load]{\includegraphics[width=0.47\textwidth]{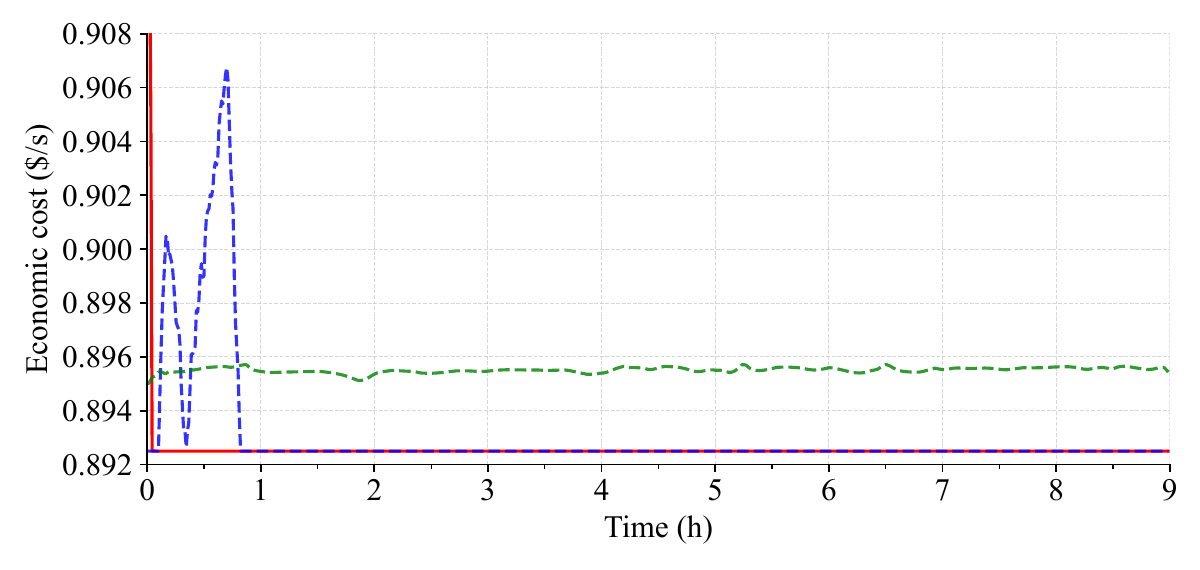}}\\
    \subfloat[\textcolor{black}{Condition 4 -- Full range operation}]{\includegraphics[width=0.47\textwidth]{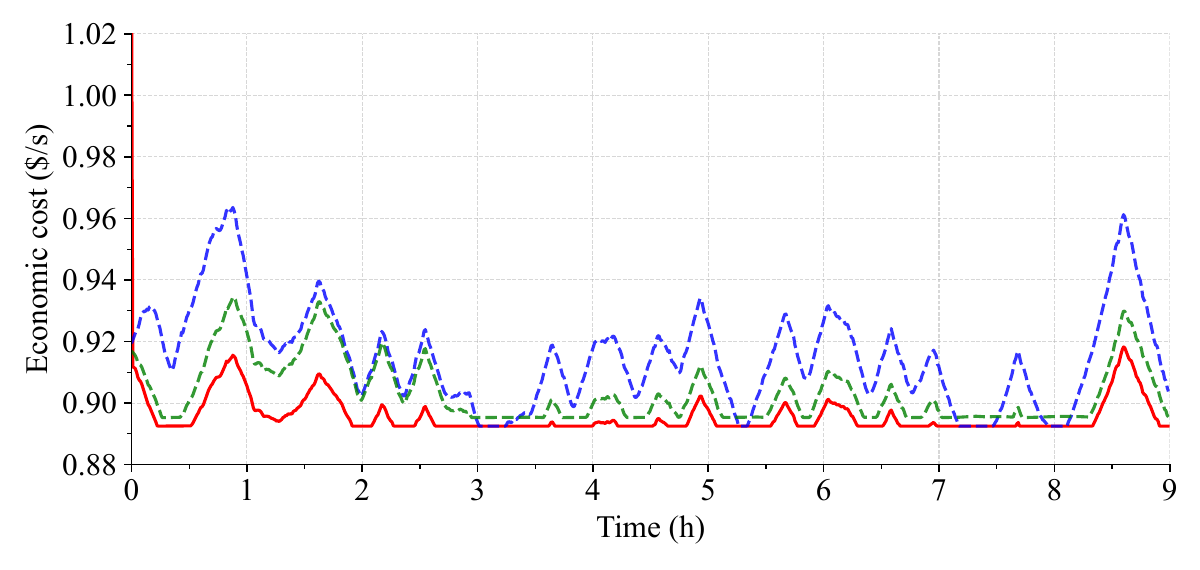}}
    \caption{\color{black}Economic costs provided by the proposed DNKO method, PI control, and SAC \cite{haarnoja2018soft}, under the four different operational conditions.}
    \label{fig:economic cost}
\end{figure}

\textcolor{black}{In our implementation, the DNNs consist of two fully connected layers, each with 128 units and ReLU as activation functions. To capture the temporal dependencies in the data, we use an LSTM network with 16 cells. The model processes a history of four time steps (\( \tau = 4 \)) to infer hidden states and make predictions. The encoded observable space is set to have a dimensionality of 40, and this ensures that the latent representation is sufficiently expressive for the task while it remains computationally efficient. To prevent overfitting and enhance generalization, the model incorporates an \( l_2 \)-norm regularization term with a coefficient of 0.1. This helps improve the robustness of the predictions for unseen data. The hyperparameters of DNKO are summarized in Table~\ref{table:hyperparameters}.}

In our evaluation, two baseline methods are considered for control performance comparisons. A PI controller \cite{seborg2016process} is designed to track an economically optimal steady-state set-point obtained by solving a static nonlinear optimization problem based on the first-principles model. At the engine load of $60\%$, the set-point is obtained as $\tilde{F}_{\text{CO}_2} = 1797.94$~kg/h and $ T_{reb}= 398.14$~K. \textcolor{black}{Using these values, the proportional ($K_p$) and integral ($K_i$) gains of the PI controller are manually tuned through trial and error to minimize tracking errors while avoiding excessive control oscillations or instability. } Additionally, we apply the soft actor-critic (SAC) algorithm \cite{haarnoja2018soft}, a model-free reinforcement learning approach, as a learning-based control baseline. As the objective of optimal process operation is to minimize the economic cost, during the training phase, SAC designed based on \cite{haarnoja2018soft}, uses the negative of the economic cost \eqref{empc:cost} as the reward function. \textcolor{black}{In addition, reward shaping is conducted to incorporate system constraints into the learning process. The $l_1$ distance to the safe set $\mathbb{Y}$ is used as a penalty term in the reward function. This penalty quantifies the magnitude of constraint violations, such as deviations of the reboiler temperature from the allowable range. The weights for the terms in the reward are tuned through trial and error.}

\begin{figure}
    \hspace{1.4cm}\includegraphics[width=0.35\textwidth]{caption-eps-converted-to.pdf}
    \subfloat[Condition 1 -- Maneuvering]{\includegraphics[width=0.47\textwidth]{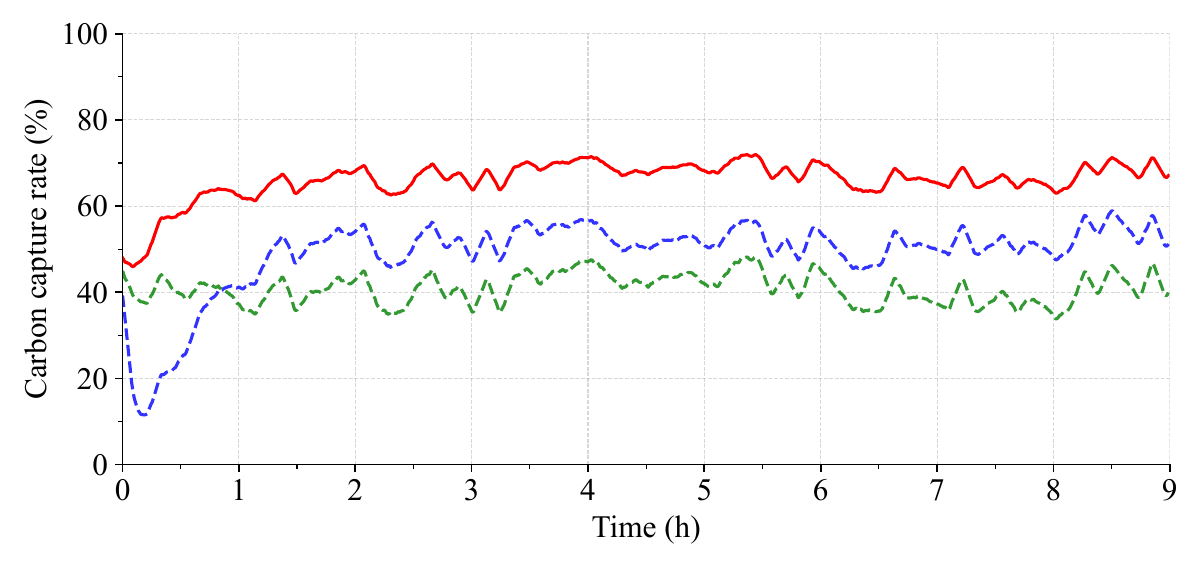}}\\
    \subfloat[Condition 2 -- Slow steaming]{\includegraphics[width=0.47\textwidth]{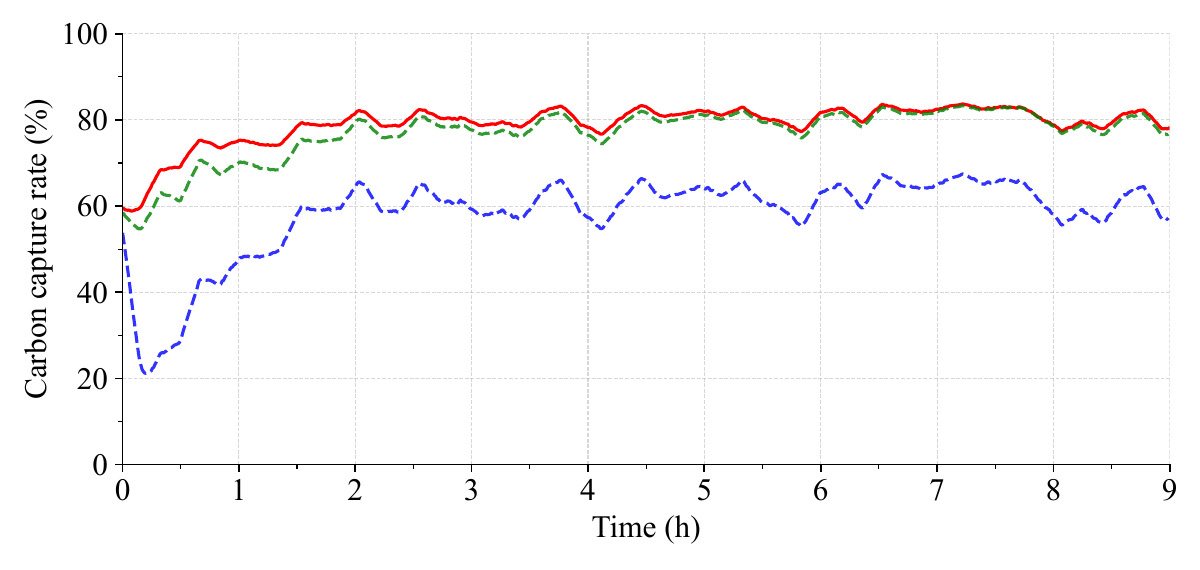}}\\
    \subfloat[Condition 3 -- Low engine load]{\includegraphics[width=0.47\textwidth]{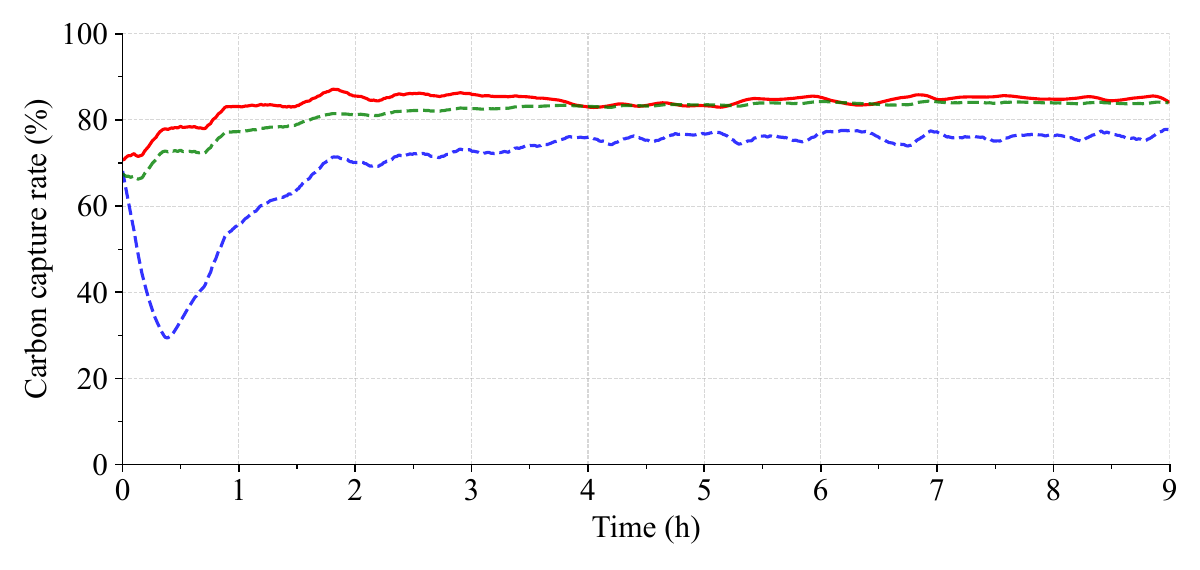}}\\
    \subfloat[\textcolor{black}{Condition 4 -- Full range operation}]{\includegraphics[width=0.47\textwidth]{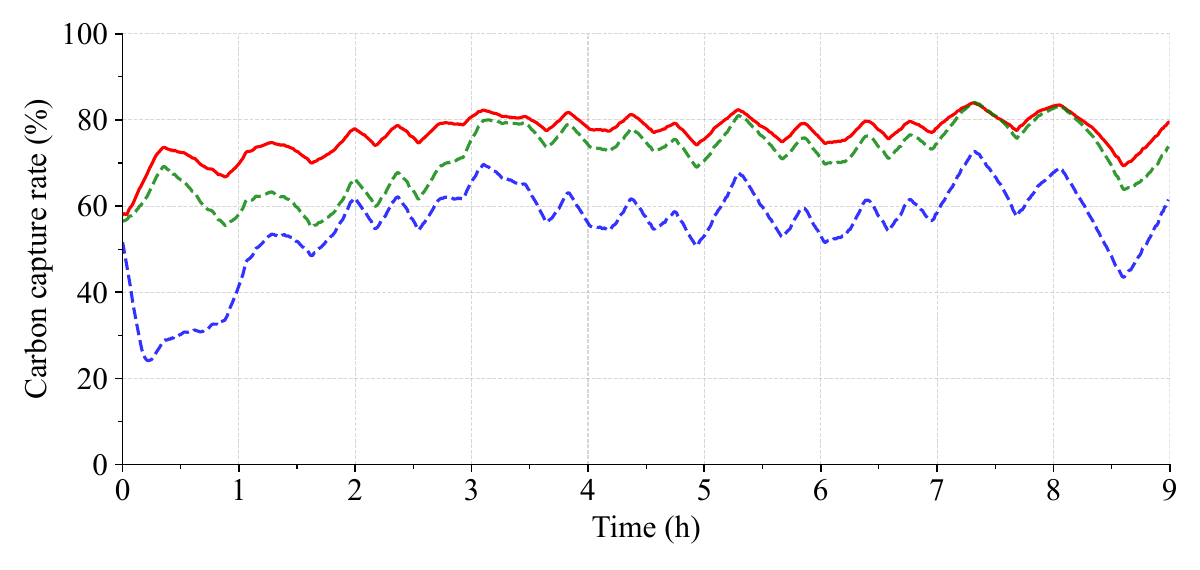}}
    \caption{\textcolor{black}{Carbon capture rates based on the proposed DNKO method, PI control, and SAC \cite{haarnoja2018soft}, under four different operational conditions.}}
    \label{fig:co2capture}
    \vspace{-1em}
\end{figure}

\begin{figure}
    \hspace{1.8cm}\includegraphics[width=0.35\textwidth]{caption-eps-converted-to.pdf}
    \includegraphics[width=\columnwidth]{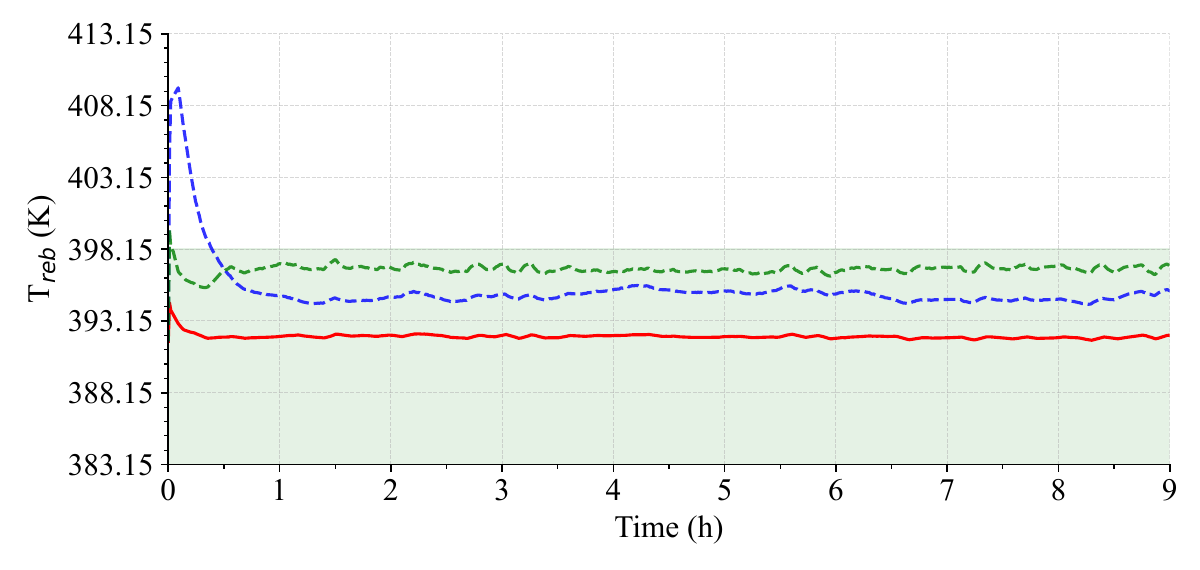}\\
    \includegraphics[width= \columnwidth]{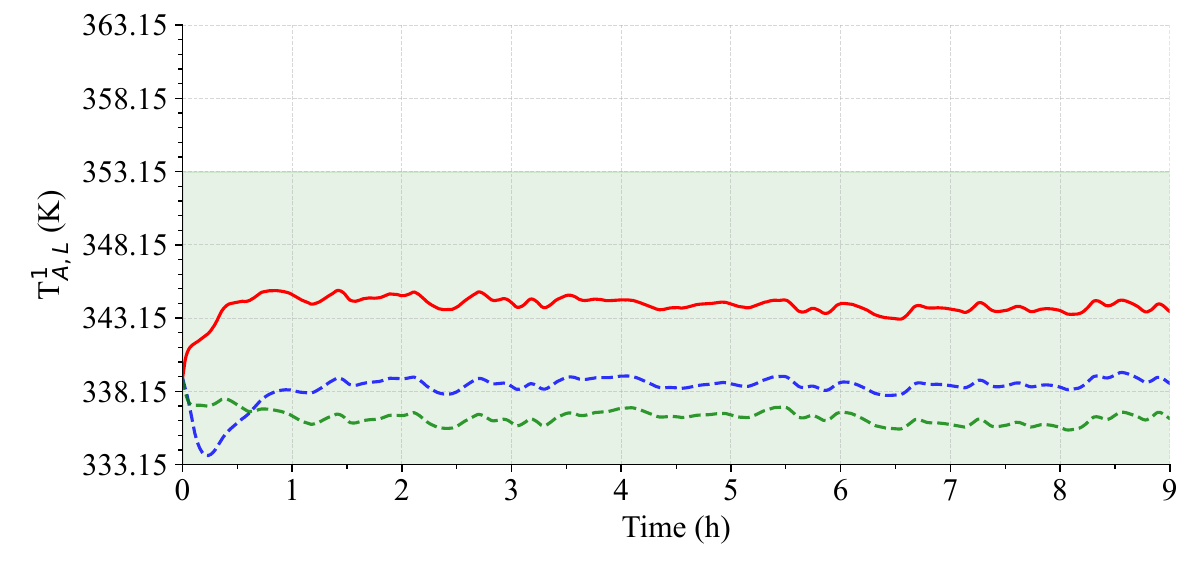}\\
    \includegraphics[width= \columnwidth]{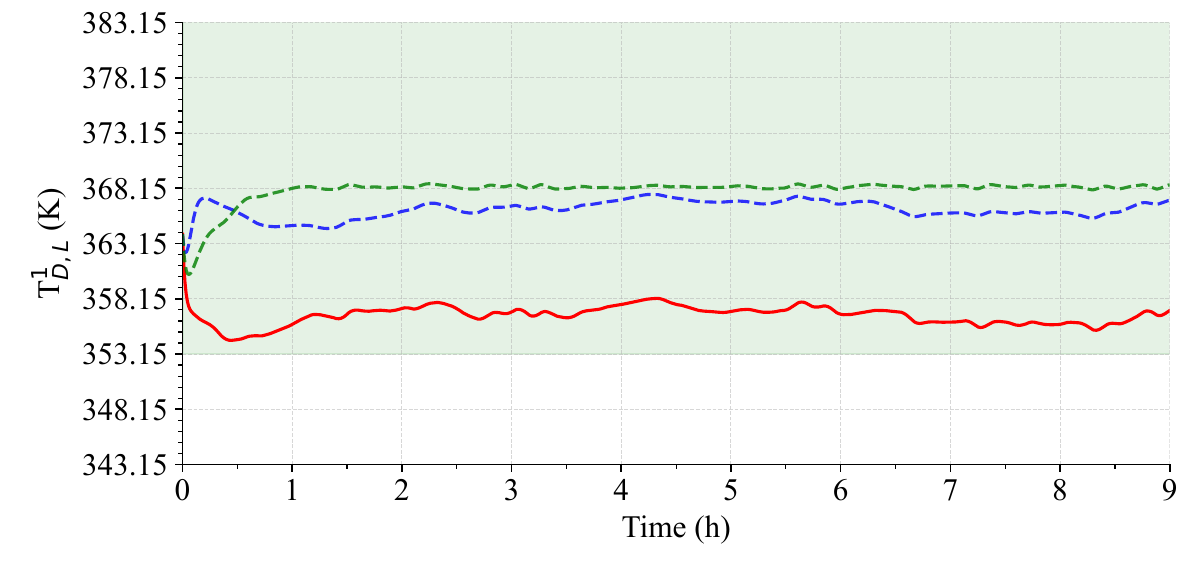}
    \caption{\textcolor{black}{Trajectories of selected controlled outputs given by the proposed DNKO method, PI control, and SAC \cite{haarnoja2018soft} under Condition 1. The feasible regions in each dimension are colored in green.}}
    \label{fig:y_high}
    \vspace{-1em}
\end{figure}

\begin{figure}
    \hspace{1.8cm}\includegraphics[width=0.35\textwidth]{caption-eps-converted-to.pdf}
    \includegraphics[width=0.48\textwidth]{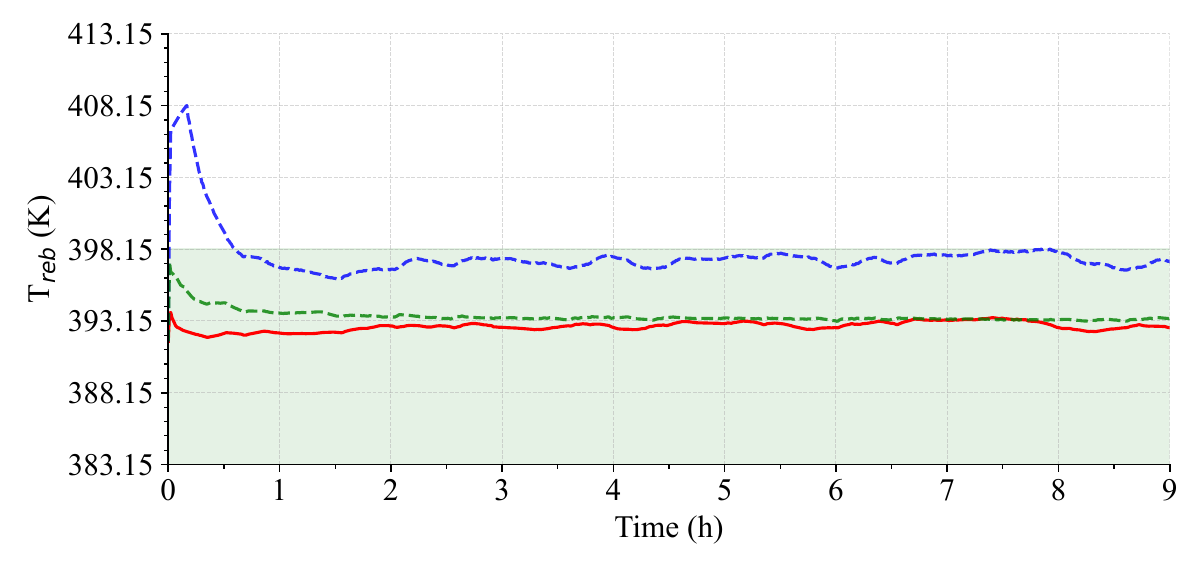}\\
    \includegraphics[width=0.48\textwidth]{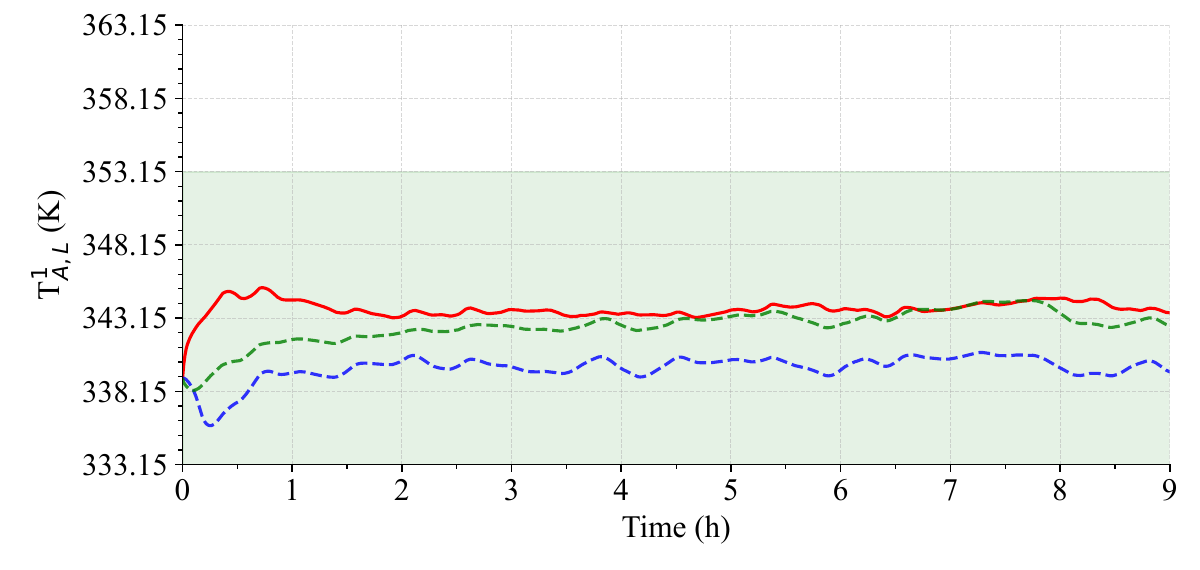}\\
    \includegraphics[width=0.48\textwidth]{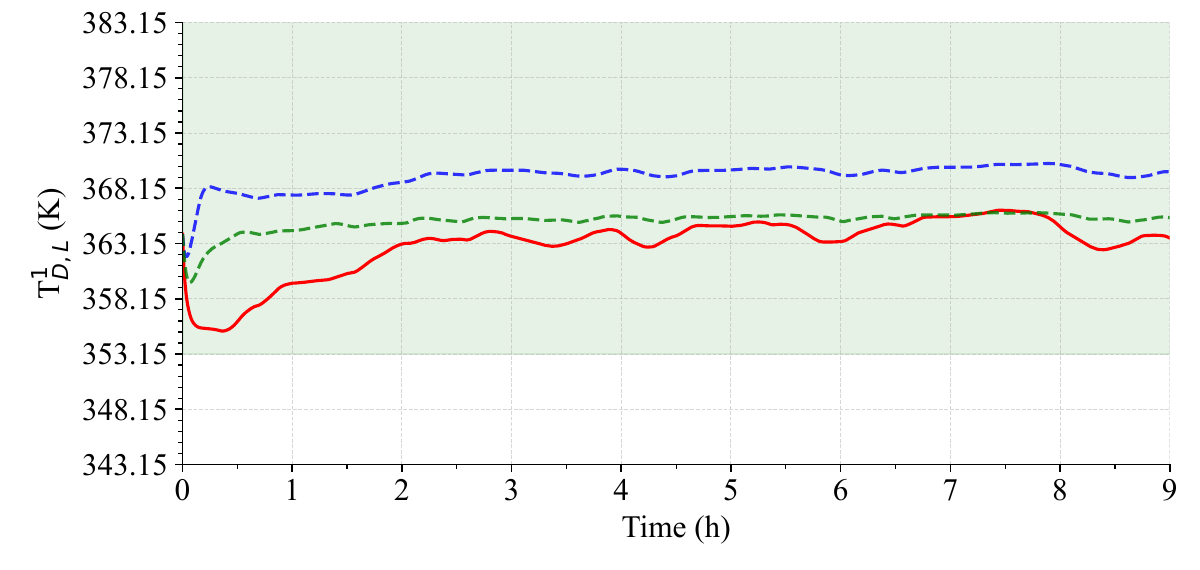}
    \caption{\textcolor{black}{Trajectories of selected controlled outputs given by the proposed DNKO method, PI control, and SAC \cite{haarnoja2018soft} under Condition 2. The feasible regions in each dimension are colored in green.}}
    \label{fig:y_middle}
    \vspace{-1em}
\end{figure}
\begin{figure}
    \hspace{1.8cm}\includegraphics[width=0.35\textwidth]{caption-eps-converted-to.pdf}
    \includegraphics[width=0.48\textwidth]{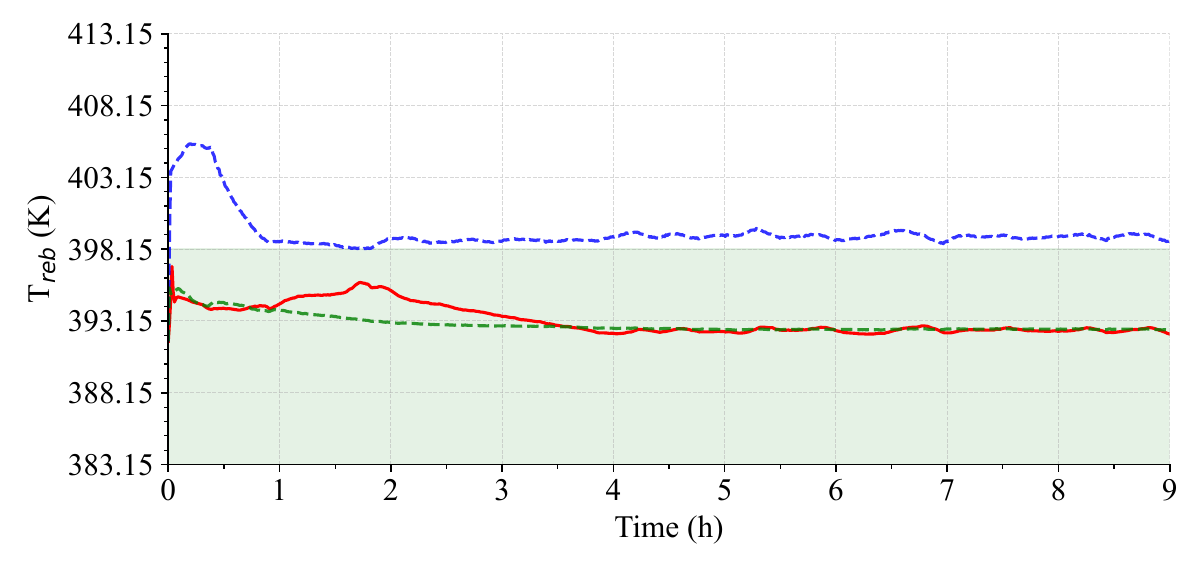}\\
    \includegraphics[width=0.48\textwidth]{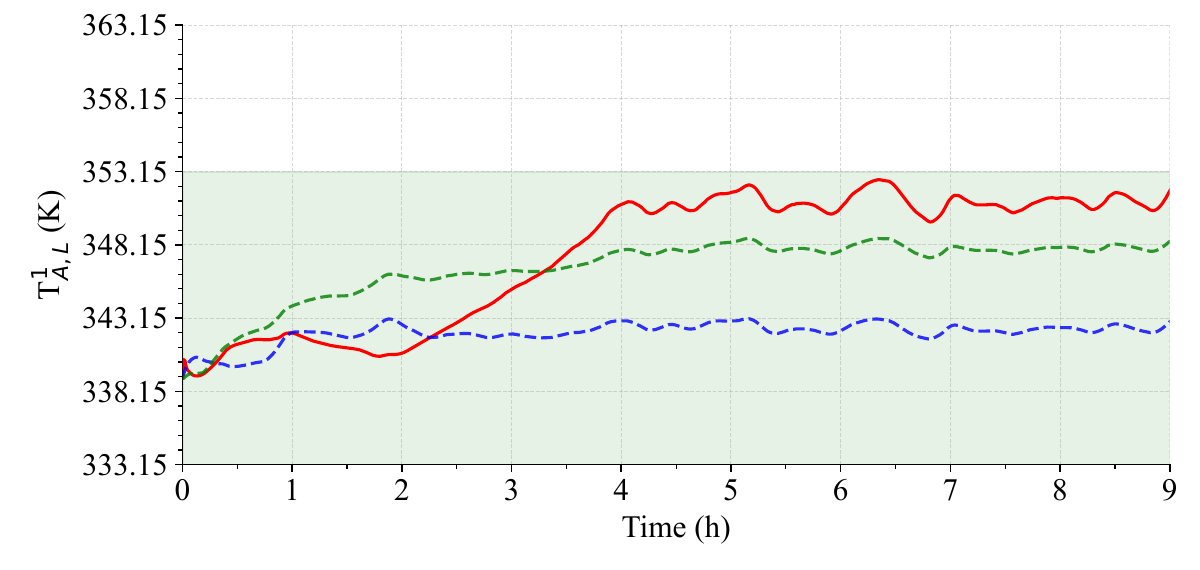}\\
    \includegraphics[width=0.48\textwidth]{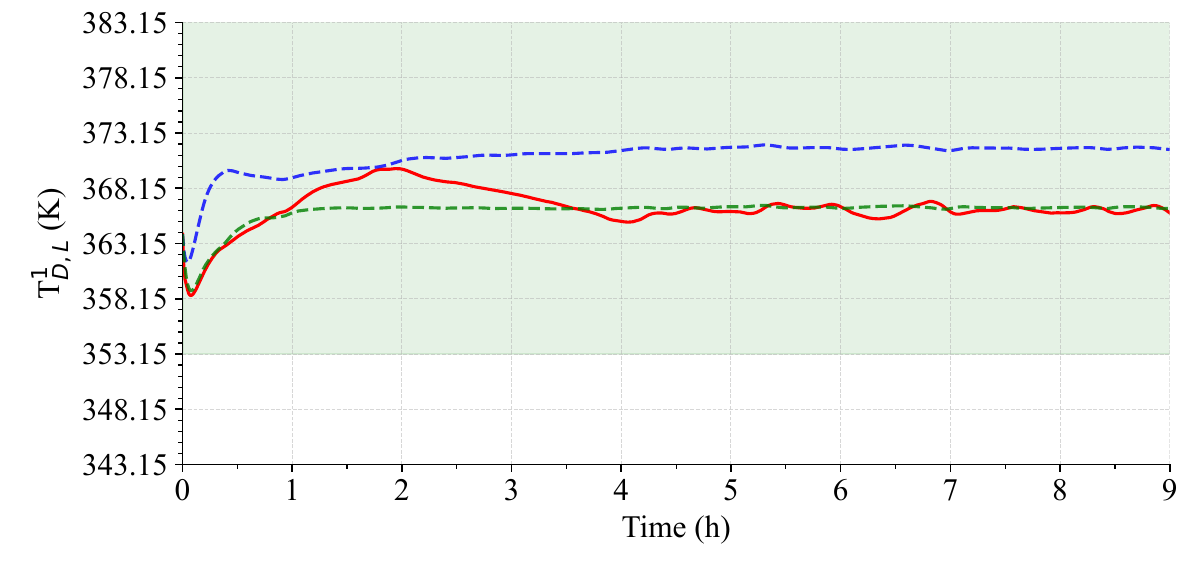}
    \caption{\textcolor{black}{System output trajectories of the proposed DNKO control method, PI control, and SAC \cite{haarnoja2018soft} under Condition 3. The feasible regions in each dimension are colored in green.}}
    \label{fig:y_low}
    \vspace{-1em}
\end{figure}

\begin{figure}
    \hspace{1.8cm}\includegraphics[width=0.35\textwidth]{caption-eps-converted-to.pdf}
    \includegraphics[width=0.48\textwidth]{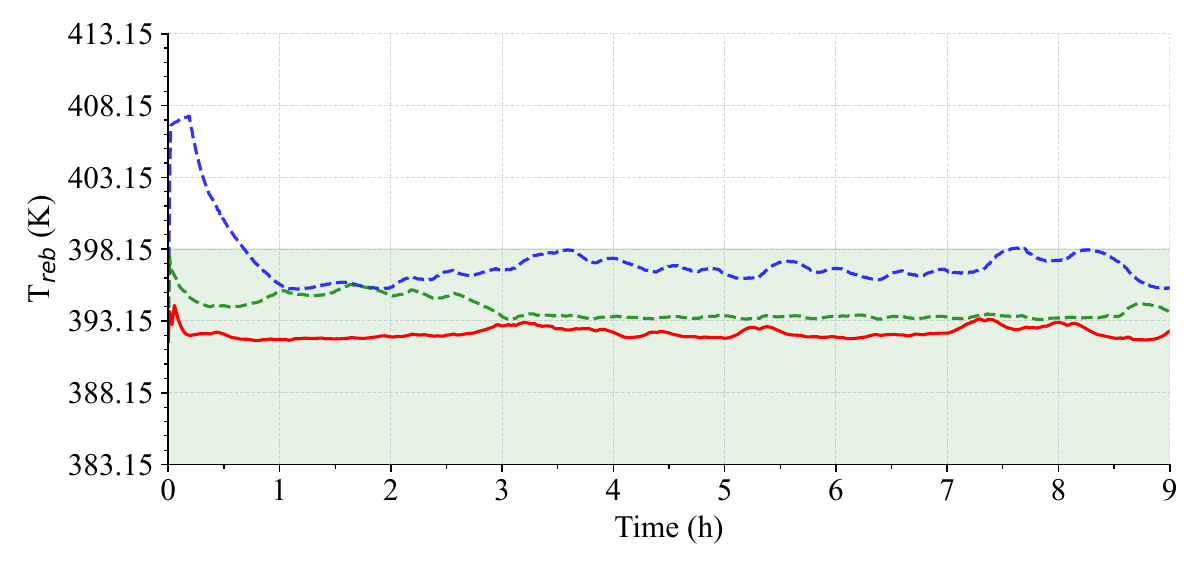}\\
    \includegraphics[width=0.48\textwidth]{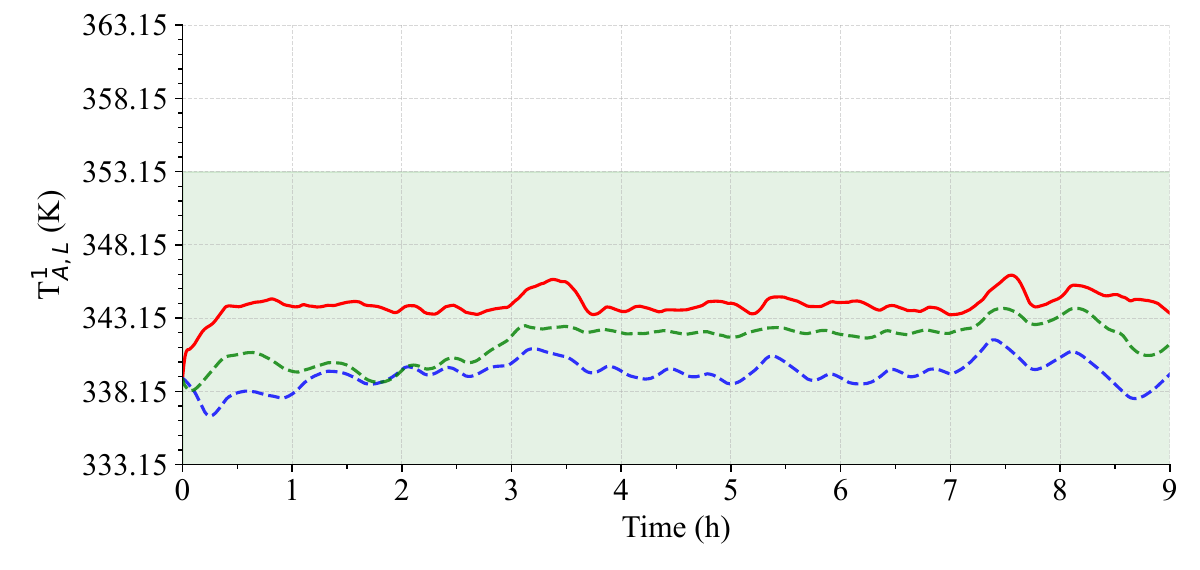}\\
    \includegraphics[width=0.48\textwidth]{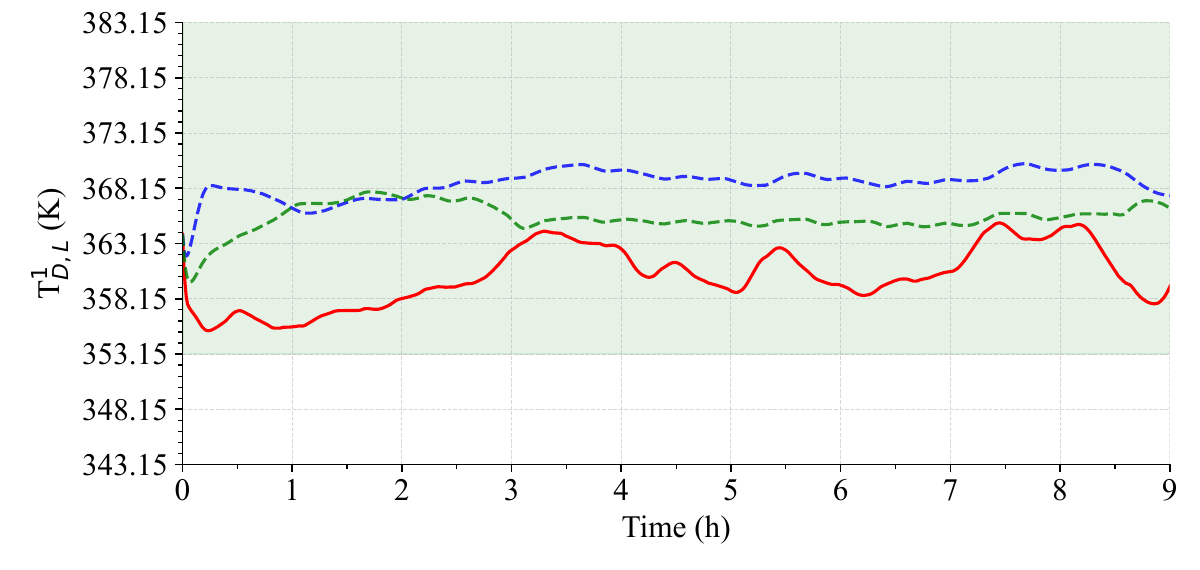}
    \caption{\textcolor{black}{System output trajectories of the proposed DNKO control method, PI control, and SAC \cite{haarnoja2018soft} under Condition 4. The feasible regions in each dimension are colored in green.}}
    \label{fig:y_vast}
    \vspace{-1em}
\end{figure}

\begin{table}
\centering
\caption{\textcolor{black}{A summary of the control performance under different operational conditions}}\vspace{2mm}
\label{table:performance summary}\renewcommand\arraystretch{1.19}
\begin{tabular}{p{2.4cm}|p{2.4cm}p{2.4cm}}
\hline
& Economic cost ($\$$)& Average carbon capture rate\\
\hline
\multicolumn{3}{l}{Condition 1 -- Maneuvering}\\
\hline
DNKO& \textcolor{black}{$\mathbf{2.970}\times 10^4$} &\textcolor{black}{\textbf{66.08}$\%$}\\
SAC (based on \cite{haarnoja2018soft}) & \textcolor{black}{${3.092}\times 10^4$} &\textcolor{black}{40.85$\%$}\\
PI & \textcolor{black}{$3.045\times 10^4$} &\textcolor{black}{48.95$\%$}\\
\hline
\multicolumn{3}{l}{Condition 2 -- Slow steaming}\\
\hline
DNKO& \textcolor{black}{$\mathbf{2.892}\times 10^4$} &\textcolor{black}{\textbf{79.09}$\%$}\\
SAC (based on \cite{haarnoja2018soft}) & \textcolor{black}{${2.902}\times 10^4$} &\textcolor{black}{77.06$\%$}\\
PI & \textcolor{black}{$2.933\times 10^4$} &\textcolor{black}{58.12$\%$}\\
\hline
\multicolumn{3}{l}{Condition 3 -- Low engine load}\\
\hline
DNKO& \textcolor{black}{$\mathbf{2.888}\times 10^4$} &\textcolor{black}{\textbf{83.89}$\%$}\\
SAC (based on \cite{haarnoja2018soft}) & \textcolor{black}{${2.898}\times 10^4$} &\textcolor{black}{81.69$\%$}\\
PI & \textcolor{black}{$2.890\times 10^4$} &\textcolor{black}{70.37$\%$}\\
\hline
\multicolumn{3}{l}{\textcolor{black}{Condition 4 -- Full range operation}}\\
\hline
DNKO& \textcolor{black}{$\mathbf{2.899}\times 10^4$} &\textcolor{black}{\textbf{76.74}$\%$}\\
SAC (based on \cite{haarnoja2018soft}) & \textcolor{black}{${2.961}\times 10^4$} &\textcolor{black}{71.22$\%$}\\
PI & \textcolor{black}{$2.924\times 10^4$} &\textcolor{black}{55.49$\%$}\\
\hline
\end{tabular}
\end{table}

\subsection{Modeling performance}

We first assess the performance of the DNKO-based modeling by training 10 DNKO models with randomly initialized parameters. Each model is trained through 1000 epochs. In each epoch, batches of 128 data points are used to update the parameters until all data is used. Fig.~\ref{fig:modeling performance} illustrates the $l_2$ norms of the prediction losses for both validation and test datasets across epochs. The validation loss converges to approximately $10^{-1}$ by the end of training, while the test loss reaches $10^{-2}$. Additionally, the shaded region in Fig.~\ref{fig:modeling performance} indicates consistent convergence over the different initial parameter settings. We note that the difference between the prediction accuracy on the validation set and the test set may further be reduced by increasing the size of the dataset. Nonetheless, the results in the subsequent subsections show that the trained model is adequate as the model basis for the controller to provide satisfactory control performance.

\subsection{Control performance}

In this subsection, we compare the control performance for the proposed DNKO EMPC method, the PI control, and the SAC method in \cite{haarnoja2018soft}. The three control methods are evaluated under the \textcolor{black}{four representative ship operational conditions}. \textcolor{black}{The engine load trajectories corresponding to the four operational conditions are presented in Fig.~\ref{fig:engine_load}.} The control performance is evaluated and compared on the basis of three performance indicators, including the economic cost, the carbon capture rate, and the satisfaction of constraints. 
The cumulative economic costs and average carbon capture rates of the proposed method and the two baselines, under different ship operational conditions, are summarized in Table~\ref{table:performance summary}. The bold numbers indicate the best performance among the three approaches. 

As shown in Fig.~\ref{fig:economic cost} and Table~\ref{table:performance summary}, DNKO achieves the lowest economic cost among the three control methods under each of the \textcolor{black}{four operational conditions}. The economic costs are higher under conditions with higher engine loads because more energy is needed to absorb an increased amount of released CO$_2$. As shown in Fig.~\ref{fig:economic cost}, the proposed DNKO-based control method can provide lower economic costs under different operational conditions. 

\textcolor{black}{
DNKO also provides the overall highest carbon capture rate among the three control methods under the four operational conditions. As shown in Fig.~\ref{fig:co2capture} and Table~\ref{table:performance summary}, the average carbon capture rate provided by DNKO remains above \textcolor{black}{66$\%$} under the four conditions. Under Condition 3, the carbon capture rate of DNKO reaches 87.12\% at peak.
}

The trajectories of the system output under the control of DNKO and the baselines under the \textcolor{black}{four} operational conditions are presented in Figs.~\ref{fig:y_high}, \ref{fig:y_middle}, \ref{fig:y_low}, and \ref{fig:y_vast}. The feasible regions in each dimension are colored in green. 

\textcolor{black}{In our evaluations, the average computation time of DNKO-based EMPC per control step is 0.69 seconds on a PC equipped with an Intel\textsuperscript{\textregistered} Core\textsuperscript{TM} i9-13900KF 3.00 GHz CPU and 64 GB RAM. With a sampling period of 40 seconds, this computation time demonstrates the proposed method’s suitability for real-time control implementation on the considered shipboard carbon capture plant. The efficiency is primarily attributed to the convex optimization formulation for the EMPC design, which can be solved using computationally efficient quadratic programming solvers, e.g., SCS \cite{ocpb:16}. }

\textcolor{black}{To further enhance the computational efficiency, several measures could be considered. 1) The Koopman operator encoding might be parallelized and can be conducted on GPUs. This would significantly reduce the overall computation time, especially for high-dimensional systems. 2) Model reduction techniques, such as Kalman-GSINDy \cite{wang2022time}, could be explored to reduce the dimensionality of the Koopman model and help improve the computational efficiency \cite{zhang2023reduced}. 3) Additional acceleration in computation could be achieved by employing specialized solvers optimized for embedded systems or leveraging warm-starting techniques to initialize the solver with solutions from the previous time step. }

\section{Conclusion and Discussion}\label{sec:conclusion}

In this work, we proposed a learning-based computationally efficient economic model predictive control framework for the safe and economically optimal operation of shipboard PCC systems. A deep neural Koopman operator (DNKO) modeling approach was proposed to predict both the economic operational cost of the PCC plant and the future information of key output variables in a latent space, using accessible partial state measurements and known inputs. By dynamically adjusting some of the parameters related to the Koopman model using the trained DNNs, the model can provide good predictions for future system outputs and key performance indices related to shipboard PCC operational performance.
Based on the established DNKO model, we formulated a convex and computationally efficient EMPC scheme. The proposed DNKO-based EMPC method was applied to the shipboard PCC system. The economic operational performance is improved as compared to two control baselines, while the operational safety is ensured and the carbon capture rate is maintained at a high level across the four representative operational conditions. 

\textcolor{black}{Physical systems, including the considered shipboard PCC considered in this work, often introduce additional complexities that cannot be fully captured by a dynamic simulator. Evaluating the proposed DNKO-based EMPC method on a lab-scale setup is an important direction of future work. 
On the other hand, incorporating data from additional operational conditions during training or fine-tuning can enhance the ability of the model to handle a broader range of operational conditions and improve the generalizability of the model.
}

\textcolor{black}{Additionally, learning the Koopman operator in a finite-dimensional observable space introduces inherent modeling uncertainties. To mitigate the impacts of these modeling uncertainties, several approaches may be considered in future research. First, integrating online adaptation \cite{nagabandi2018learning} into the DNKO-based framework is promising for dynamically mitigating modeling errors. Online learning may also the model to update its parameters in real-time by leveraging new data collected from system operations \cite{nagabandi2018learning}. 
Second, employing tube-based MPC \cite{langson2004robust,zhang2022robust} is one of the promising ways to ensure constraint satisfaction in the presence of modeling uncertainties. 
Third, robust EMPC \cite{bayer2016robust,schwenkel2020robust} can incorporate uncertainties into the optimization problem; this way, the robustness of control performance can be enhanced in the presence of model uncertainties. Integrating the concept of robust EMPC (e.g., \cite{bayer2016robust,schwenkel2020robust}) with the proposed modeling framework may help provide economic control performance guarantees for nonlinear systems with uncertainties. }

\bibliography{references,Control}
\bibliographystyle{IEEEtran}

\end{document}